\pdfoutput=1
\documentclass{siamart1116}

\ifpdf
  \DeclareGraphicsExtensions{.eps,.pdf,.png,.jpg}
\else
  \DeclareGraphicsExtensions{.eps}
\fi

\numberwithin{theorem}{section}

\newcommand{\TheTitle}{Dynamic Looping of a Free-Draining Polymer} 
\newcommand{\TheAuthors}{Felix X.-F. Ye, Panos Stinis  and Hong Qian}

\headers{\TheTitle}{\TheAuthors}

\title{{\TheTitle}\thanks{Submitted to the editors April 24th, 2017.
}}

\author{
 Felix X.-F. Ye\thanks{University of Washington, Department of Applied Mathematics, Seattle, WA.}
  \and
 Panos Stinis\thanks{Pacific Northwest National Laboratory, Richland, WA.}
  \and
  Hong Qian\footnotemark[2]
}

\usepackage{amsopn}

\ifpdf
\hypersetup{
  pdftitle={\TheTitle},
  pdfauthor={\TheAuthors}
}
\fi

\usepackage{graphicx}
\usepackage{subfig}
\usepackage{amssymb}
\usepackage{amsmath}
\usepackage{amsbsy}
\usepackage{cleveref}
\usepackage{cancel}
\usepackage{bbm}
\usepackage[export]{adjustbox}
\usepackage{multirow}
\DeclareMathOperator\erf{erf}

\def\mR{{\bf R}}
\def\mF{{\bf F}}
\def\mX{{\bf X}}

\def\mI{{\bf I}}

\def\mr{{\bf r}}

\def\rd{{\rm d}}
\def\vf {{\bf f}}

\begin{document}

\maketitle

\begin{abstract}
We revisit the celebrated Wilemski-Fixman (WF) treatment 
for the looping time of a free-draining polymer.  The WF theory
introduces a sink term into the Fokker-Planck equation for 
the $3(N+1)$-dimensional Ornstein-Uhlenbeck process of 
the polymer dynamics, which 
accounts for the appropriate boundary condition due to the
formation of a loop. The assumption for WF theory is considerably relaxed. A perturbation method approach is 
developed that justifies and generalizes the previous
results using either a Delta sink or a Heaviside sink.  
For both types of sinks, we show that under the condition 
of a small dimensionless $\epsilon$, the ratio of 
capture radius to the Kuhn length, we are able to systematically
produce all known analytical and asymptotic results obtained 
by other methods.  This includes most notably the transition 
regime between the $N^2$ scaling of Doi, and $N\sqrt{N}/\epsilon$ scaling of
Szabo, Schulten, and Schulten.  The mathematical
issue at play is the non-uniform convergence of $\epsilon\to 0$
and $N\to\infty$, the latter being an inherent part of the theory
of a Gaussian polymer.   Our analysis yields a novel term in the analytical expression for the looping time with small $\epsilon$, 
which was previously unknown.  Monte Carlo numerical simulations 
corroborate the analytical findings. The systematic
method developed here can be applied to other systems modeled 
by multi-dimensional Smoluchowski equations.
\end{abstract}



\section{Introduction}
The theory of polymer dynamics is one of the most 
successful stochastic-process models in chemical science
\cite{flory1989polymer,DoiEdwards1988}.  Polymer dynamics in
an aqueous solution is naturally stochastic.  A 
polymer molecule dissolved in the so-called theta solvent,
known also as free-draining, can be 
mathematically represented in terms of a 
multi-dimensional Ornstein-Uhlenbeck (OU) process.  
While the general theory of an OU process 
is well developed (see \cite{Qian-OU} and the references cited
within), explicitly analytical results on the kinetics of the
formation of a end-to-end loop are still highly sought after in
theoretical chemistry and biochemistry. 
The problem is essentially a 
perturbation of a linear operator \cite{kato-book,ward1993,ward2011}.

In the literature, ``polymer looping'' has been a classic problem with important biological applications.
It is an essential step in various intra-cellular processes, including gene regulation, DNA replication, protein and RNA folding \cite{DNA-looping-review1, DNA-looping-review2, nature1986}. In some biological systems, two reacting molecules may bind distal target sites along a single polymer chain and a biological function requires the polymer to ``spontaneously'' form a loop structure that brings the two in contact \cite{levene2013}. This polymer-loop-mediated phenomena is due to thermal fluctuations and the loop formation consists of two steps:
the ``encounter'' of the two ends of the polymer defined by a small
capture distance, followed by a bond formation between the two 
ends when they are within the capture radius.  When the 
time scale of the former is much longer than the latter, the
loop formation is called diffusion controlled: The
diffusive Brownian motion of the chain is then the rate determining step. 

	As an in-depth study of a multi-dimensional stochastic process,
the problem was first discussed by Jacobson and Stockmayer in 1950
even before the dynamic Rouse model \cite{jacobson1950,rouse1953}.
They formulated the free energy cost of looping in terms of an equilibrium distribution. It is quantified as the ratio of respective equilibrium constants for intermolecular and intramolecular synapsis reactions when studying the polymer melts, which is called the J factor. The J factor is directly proportional to the statistical fraction of polymer conformations within the capture radius $a$, which is the ratio of the rate constants of the cyclization reaction, $k_{\text{coil}\rightarrow \text{loop}}/k_{\text{loop}\rightarrow \text{coil}}$. 
The J factor is mainly discussed in the more complicated case where the effect of bending and twisting rigidity is taken into account and so the polymer is no longer flexible \cite{shore-baldwin}.  
The rate $k_{\text{coil}\rightarrow \text{loop}}$ is the inverse of the mean time for the two ends of a polymer to meet within the capture radius $a$ (or simply looping time $\tau$) from some configurations.

The study of the kinetics of loop formation is also a classical problem in polymer dynamics. It has led to intense theoretical and numerical research, even for the simplest case, of a Rouse chain. 
Though seemingly simple, the problem is actually challenging \cite{Szabo1996}. The mathematical formulation of the kinetics of this looping process involves a non-simple boundary-value problem of a linear PDE in a high dimensional space. 

There were two major theoretical 
approaches in the early stage of the polymer looping investigation;
they led to seemingly contradicting results. First, Wilemski and Fixman (WF) \cite{WilemskiFixman1, WilemskiFixman2} and Doi \cite{Doi1975} estimated that the looping 
time for long polymer chains scales as  $\sim N^2$, where $N$ is the number of monomers. Surprisingly, according to this result, the looping time does not depend on the capture radius $a$ (more details of this method will appear in later sections). On the other hand, Szabo, Schulten and Schulten (SSS) \cite{Szabo1980} estimated the looping time scales as $\sim N^{3/2}/a$. They approximated the dynamics of the end-to-end distance of the polymer as a single Brownian particle diffusing in a potential of mean force (this theory will be discussed in \Cref{Markovian}).  Recent numerical simulations, however, indicated these time scales are showing up in different time regimes but the prediction from WF is more accurate than that from SSS theory \cite{Chen2005, Szabo1996}.  In fact, more recent theoretical advances show that the looping time may follow a mixed scaling law with $\tau\approx c_1 N\sqrt{N}/a+c_2 N^2$. Toan {\it et al} \cite{Toan2008} proposed an effective 
space-dependent diffusion coefficient in SSS theory, such that both theories could be unified into one and the mixed scaling law was revealed. However, it has been difficult to verify whether 
this space dependency is really a mathematical approximation of a 
multi-dimensional OU process, or a different mathematical model all together.
Very recently Amitai {\it et al} \cite{Amitai2012-Flexible} also discovered a similar scaling law, based on the expansion of the eigenvalues of the Fokker-Planck equation in the limit of small capture radius. The first order perturbation in $a$ of the largest eigenvalue matches the SSS result and the next order perturbation roughly scales as $N^2$. But their theory works well only when $N$ is relatively small ($N <64$), and the contribution from other eigenvalues is unclear. Gu\`erin {\it et al} \cite{nature2016} and B\`enichou {\it et al} \cite{ Renewal2015} adapted the renewal equation method on the non-Markovian process of the end-to-end vector and reviewed other previous approaches. The numerical result they obtained is better than previous theories but an analytic scaling law is currently out of reach.   


Within the OU process framework and 
in terms of analytical results, the mean first passage time is a
solution to a boundary value problem associated with a 
high dimensional backward equation.   Significant progress has been made in recent years on the
asymptotic of passage-time problem associated with
three-dimensional diffusion processes with a small exit boundary \cite{ward2011,bressloff-rmp,tomchou}, crossing an unstable limit-cycle barrier \cite{daoduc}, and rigorous lower 
bound on the density of the passage-time of the OU process \cite{peterthomas}.   The WF theory
distinguishes itself from these other works in dealing with 
a high-dimensional OU process, which is equivalent to a
low-dimensional non-Markovian process.
With this in mind, the WF theory is currently the only analytically feasible approximation for the non-Markovian process of the end-to-end vector.

 In the more than 40 years since the WF theory was first proposed, many efforts have been made to improve this theory numerically and theoretically. 
A comprehensive theoretical discussion is given in \cite{Weiss}. In particular, WF theory was found to agree well with simulation results for the case of small capture radius. However, no explicit analytical expression was provided and instead numerical integration was employed. Pastor {\it et al} \cite{Szabo1996} observed that this integral can be evaluated in the small time range in the limit of small capture radius $a$ and one finds the same limit as SSS. Moreover, they claimed the looping time must combine both kinds of behavior, $N^{3/2}/a$ and $N^2$. This shows the potential to extract the mixing scaling law from WF theory. 
Sokolov \cite{Sokolov2003} and Likthman {\it et al} \cite{Likthman2006} further improved the WF theory to accommodate different initial conditions and used iterative methods to find exact solutions (for both cases numerical calculations were needed).

This paper is organized as follows. To make the presentation
self contained, in \Cref{Rouse} we review the Rouse model and derive the end-to-end vector as the summation of independent, non-identical OU processes. We express the looping time in terms of a 
boundary-value problem for a $3(N+1)$-dimensional Fokker-Planck equation. In \Cref{Markovian}, we review the Markovian approximation for the end-to-end vector and show that SSS theory corresponds to the first order perturbation of the largest eigenvalue for the Fokker-Planck equation. In \Cref{WF}, we discuss the WF theory rigorously and articulate the assumptions behind it. Specifically, we show that the assumption for the Delta sink is weaker than the Heaviside sink, so that the Delta sink gives better approximation results. The integrand appearing in WF theory in the expression for the looping time is not a genuine survival probability and we give a probabilistic interpretation for the Heaviside sink. The theory presented in this section is much more general and could be applied to other types of boundary-value problems.  
In \Cref{perturb-section}, we extract the postulated mixed scaling laws directly from WF theory in the limit of small dimensionless $\epsilon$, the ratio of capture radius $a$ to the Kuhn length $b$, with the help of perturbation methods. The mixed scaling law we find includes an extra term $N\sqrt{N}$ which will be dominated by $N^2$ term as $N\rightarrow +\infty$. In \Cref{numerical-simulation}, we use Monte Carlo simulations to estimate the looping time for various $N$ and the capture radius $\epsilon$. In general, WF theory overestimates the looping time but agrees very well with the simulation in the small $\epsilon$ regime (irrespective of the value of $N$). We also verify the mixed scaling law numerically through regression. Last but not least, the survival probability function is numerically obtained by deconvolution and for small $\epsilon$, it is a single exponential function and agrees remarkably well with numerical simulation.

\section{Rouse model} \label{Rouse}
The Rouse model is the simplest flexible polymer model
which uses beads and springs
to represent the polymer chain in a viscous fluid. It assumes beads have no excluded volume and no hydrodynamic interactions among them. Actually, the notion of a theta solvent is defined as an exact cancellation between these two opposing effects.
Two adjacent beads are connected with a harmonic spring with the same spring constant $k$ \cite{DoiEdwards1988, klapper1998} and in total there are $N+1$ beads and $N$ springs. The potential energy of the 
entire mechanical system, $H(\vec{\mR})$, is given by 
$H(\vec{\mR})=\frac{1}{2}k\sum_{n=1}^N(\mR_n-\mR_{n-1})^2$,
where $\mR_n$ is the position of $n$th bead in three-dimensional space. Let $b$ denote the Kuhn segment length or effective bond length, that is defined as the mean square length of the bond. Then, if $k=3/b^2$ in $k_BT$ units, the Boltzmann distribution of the conformation and the end-to-end vector distribution will be the same as the ones for the Gaussian chain.  In addition to the Hookean force from its connecting springs, a bead experiences a frictional force when it moves. Each bead is assumed to
have the same friction coefficient $\zeta$.  Inertia is negligible since the motion is overdamped. The diffusion coefficient of a monomer is $D_0=1/\zeta$ in $k_BT$ units as dictated by the Einstein relation. The dynamics are described by Langevin equations \cref{langevin-1} with Gaussian white noise \cite{VanKampen}, 
 \begin{equation}\label{langevin-1}
 \frac{\rd \mR_i}{\rd t}=-D_0\nabla_{\mR_i}H(\mR)+\vf_i \ \ i=0,1,\dots,N,\ \ \langle \vf_n(t)\vf_m(t')\rangle = 2D_0\mI\delta_{nm}\delta_{tt'}
\end{equation}
Through a normal mode transformation (\ref{normal-mode}), the Langevin equations \cref{langevin-2} for the modes $\mX_p$ are decoupled and each mode evolves independently but not identically to the others. 
\begin{align}\nonumber
   &\theta_p=\frac{p\pi}{2(N+1)}  ,\ \gamma_p = 12\sin^2( \theta_p), \\ \label{normal-mode}
 &\mX_p=\frac{1}{N+1}\sum_{n=0}^N \mR_n \cos\Big((2n+1)\theta_p \Big),  \ \ \mR_n=\mX_0+2\sum_{p=1}^N \mX_p \cos\Big((2n+1)\theta_p \Big), \\ \label{langevin-2}
 &\frac{\rd \mX_p}{\rd t}=-\gamma_p\mX_p+\mF_p,\ \langle \mF_p(t)\mF_q(t')\rangle= \frac{1}{N+1}\mI\delta_{pq}\delta_{tt'}\  \text{for}\ p+q>0  
   \end{align}
Here $\gamma_p$ are eigenvalues of the Laplacian matrix of the Rouse chain. 
The zeroth mode of the chain represents the dynamics of the center of mass for the chain which is Brownian motion. The dynamics is also called a reversible Ornstein-Uhlenbeck process \cite{Qian-OU} and its equilibrium distribution is the Boltzmann distribution $P_e(\mX)$ in \cref{Boltzmann} \cite{qian-jmb}. 
   \begin{align} \label{Boltzmann} 
 &P_e(\mX)= \prod_{p\ \text{odd}}\Big(\frac{(N+1)\gamma_p}{\pi}\Big)^{3/2}\exp(-(N+1)\gamma_p\mX_p^2)
\end{align}
Nondimensionalize the system with the Kuhn length $b$ as the characteristic length and $b^2/D_0$ as the characteristic time, one can define a dimensionless parameter as the ratio of capture radius to the Kuhn length, $\epsilon=a/b$. It specifies the capture radius in the unit of the Kuhn length so we will call $\epsilon$ as capture radius as well. 
From now on, all quantities and equations are nondimensionalized unless otherwise specified. 

The end-to-end vector $\mR_{ee}$ can be expressed as the linear combination of the odd order terms of $\mX_p$ with coefficient $c_p$ (see \Cref{end-to-end-vector}). All even order modes do not contribute if the chain has a homogenous spring constant. The equilibrium distribution of the end-to-end vector is also a Gaussian with variance $N $. Then the J factor has the simple expression as scaling of $N$, $J\propto (3/2\pi N)^{3/2}$.
\begin{align}\label{end-to-end-vector}
&\mR_{ee}=\mR_0-\mR_N=\sum_{p\ \text{odd}} c_p\mX_p, \ c_p=4\cos(\theta_p),\\
 & P_{\mr}(\mR_{ee})=\Big(\frac{3}{2\pi N }\Big)^{3/2}\exp\Big(-\frac{3\mR_{ee}^2}{2N }\Big)
\end{align}
The relaxation time $\tau_p$ for the odd modes $\mX_p$ is $\tau_p=1/\gamma_p$. Specifically, $\tau_1$ is the largest, $\tau_1\approx \frac{(N+1)^2}{3 \pi^2}$. So the relaxation time for the end-to-end vector $\mR_{ee}$ is dominated by $\tau_1$. 
 
 The corresponding Fokker-Planck equation for the Langevin equation \cref{langevin-2} describes the time evolution of the probability density $P(\mX,t)$ \cite{Gardiner}. This linear operator is denoted as $L_F.$ The Green function of the Fokker-Planck equation is the probability propagator,  $G(\mX, t |\mX^0, 0),$ and is a Gaussian function. 
 \begin{align}  \label{F-P}
 \frac{\partial P(\mX, t)}{\partial t} &=&\nabla\cdot \Big( \Gamma\mX P(\mX, t)\Big)+\frac{1}{2(N+1)}\nabla\cdot\Big(\nabla P(\mX,t)\Big)=L_FP(\mX, t),
 \end{align}
in which $\Gamma$ is a diagonal matrix with elements
$\gamma_p$.
But the dynamics of the end-to-end vector $\mR_{ee}$, being a projection of a $N$-dimensional OU process, is {\it non-Markovian}. One could sample from Boltzmann (canonical) distribution polymer configurations with fixed $\mR_{ee}=\mr_0$ and study the average end-to-end vector dynamics. These dynamics follow a non-Markovian Gaussian process with the conditional probability $T(\mr, t |\mr_0,0)$. The probability at time $t$ on location $\mr$ is also w.r.t. the canonical ensemble. It is determined by the time correlation function of the end-to-end vector, $\phi(t)$, which is the summation of the correlation functions of the odd modes. The conditional probability $T(\mr, t |\mr_0,0)$ looks like a probability propagator, but it doesn't satisfy the Kolmogorov-Chapman equation, i.e, $\int T(\mr, t |\mr', t')T(\mr', t' |\mr_0, 0 )\rd \mr' \ne T(\mr, t |\mr_0,0)$, for $t>t'>0$.
\begin{eqnarray} \label{correlation}
 &\phi(t)=\frac{\langle \mR_{ee}(t) \mR_{ee}(0)\rangle}{\langle \mR_{ee}^2\rangle} 
 =\frac{2}{N(N+1)}\sum_{p\ \text{odd}}\frac{1}{\tan^2(\theta_p)}\exp(-\gamma_p t)  \\ \label{Greens}
  & T(\mr,t | \mr_0,0)=\Big(\frac{3}{2\pi N (1-\phi^2(t))}\Big)^{3/2}\exp\Big(-\frac{3}{2N }\frac{(\mr-\phi(t)\mr_0)^2}{1-\phi^2(t)}\Big)
  \end{eqnarray}
Once the two ends of the chain get close (within the capture distance $\epsilon$), the reaction may start. So this reactive surface is described as a tubular neighborhood $S_\epsilon=\{\mX: \|\mR_{ee}\|=\epsilon\}$. In the ideal case, every time the two ends get closer than $\epsilon $, the reaction happens and this corresponds to the absorbing boundary condition $P(\mX, t)=0$ on the surface $S_\epsilon$. 
Once the reaction term is imposed, the modes are no longer independent. 

In the literature, when addressing the looping problem, it is assumed that the initial distribution is the Boltzmann distribution $ P_e(\mX)$. But the initial condition doesn't fulfill the boundary condition. Physically all configurations that are in a looped state react immediately, so the initial condition becomes the normalized Boltzmann distribution outside the reactive surface and zero inside the surface. The difference between these two initial conditions are negligible since the probability inside the surface is very small compared with outside.  
It is worth noting that the polymer chain sampled from the Boltzmann distribution doesn't imply all internal modes of the chain are fully relaxed. 

If the PDE  \cref{F-P} with the boundary condition specified could be solved, one could integrate the solution $P(\mX,t)$ over the configuration space $\mX$ and get the probability that the chain hasn't formed a loop at time $t$, the survival probability $\Sigma(t)$. 
Then $1-\Sigma(t)$ is the cumulative probability distribution
for the random time of capturing, and the mean
looping time $\tau$ is the time integral of survival probability through the time domain with some mild assumptions, which can be fully determined. $\tau$ is a function of the number of beads $N$ and the dimensionless capture radius $\epsilon$.  One intuitively expects
that $\tau\to\infty$ as $N\to\infty$, or $\epsilon\to 0$.  

\section{Markovian approximation} \label{Markovian}
A simple approach is to consider that the two ends of the polymer behave as if they were diffusing apart from the rest of the chain, such that no memory effect is expected. We can project all these internal modes of the polymer into a 1D single variable, the end-to-end distance $r=\|\mR_{ee}\|$ (after angular components are averaged).  
This is the theory first developed by Szabo, Schulten and Schulten \cite{Szabo1980}. Suppose that the end-to-end distance obeys a reduced Langevin equation of the form with diffusion coefficient D in the unit of $D_0$,
\begin{equation}
\frac{\rd r}{\rd t}=-D\frac{\rd U(r)}{\rd r}+\xi(t),\ \langle \xi(t)\xi(t')\rangle= 2 D\delta(t-t'), 
\label{Langevin-SSS}
\end{equation}
where the effective potential $U(r)$ of the mean field is given by $U(r)=-\ln P_e(r)$, where $P_e(r)=4\pi r^2\Big(\frac{3}{2\pi N }\Big)^{3/2}\exp\Big(-\frac{3r^2}{2N }\Big)$. 
One can calculate the mean first passage time $\tau(r)$ starting from the fixed end-to-end distance $r$ and integrate this space dependent function w.r.t. the equilibrium distribution to get the looping time $\tau$. 

The dynamics can be considered those of a single particle diffusing in the effective potential. The looping problem becomes a standard first passage time problem which can be addressed by solving the adjoint operator equation,
$D\Big(\frac{\rd^2 \tau(r)}{\rd r^2}-\frac{\rd U}{\rd r}\frac{\rd}{\rd r}\tau(r)\Big)=-1,$  
with boundary condition $\tau(\epsilon )=0$. One can assume a reflective boundary condition when the chain is fully stretched, i.e. $\frac{\rd}{\rd r}\tau(N )=0.$ In fact, the difference with the natural boundary condition here is negligible since the probability of the chain reaching that far is very small for the flexible chain. It is easy to show the solution of the adjoint equation is 
\begin{equation}
\tau(r)=\int_\epsilon^r\frac{1}{D\exp(-U(x))}\rd x\int_x^\infty \exp(-U(y))\rd y
\end{equation}
Averaging over the equilibrium distribution of the end-to-end distance and find, 
\begin{equation}\label{SSS}
\tau=\int_\epsilon^{\infty}\tau(r)P_e(r)\rd r =\int_\epsilon^\infty \frac{1}{DP_e(x)}\rd x\Big(\int_x^\infty P_e(y)\rd y\Big)^2
\end{equation}
\Cref{SSS} is known as the SSS theory of polymer looping. It can be simplified through the Laplace method if we assume that the potential barrier at $r=\epsilon$ is very high. In SSS theory, the diffusion coefficient $D$ is chosen as the relative diffusion coefficient of the two end monomers, $D=2.$ (in \Cref{perturb-section}, it is shown that this choice of diffusion coefficient is related to a short time approximation for the end-to-end correlation function $\phi(t)$). For the Rouse model, the looping time becomes 
\begin{align}\label{SSS-approx}
\tau_{\text{SSS}}\approx \frac{1}{8\pi \epsilon  P_e(0)}=\frac{\sqrt{\pi}}{6\sqrt{6}}\frac{N\sqrt{N}}{\epsilon}
\end{align}
Specifically, $P_e(0)$ is exactly proportional to the J factor of the Gaussian chain. Another way to derive the SSS result is to coarse grain the polymer chain to a dumbbell model with the effective spring constant $k_{\text{eff}}=3/N.$ 

It turns out that the equilibrium assumption about the chain is not an accurate one. Not all internal modes of the polymer chain have relaxed before the looping process starts. So when $\tau_1\gg \tau_{\text{SSS}}$, the effective potential is no longer time independent. The motion of all internal modes has be taken into consideration in the problem of looping. If $\epsilon\ll 1/\sqrt{N} $, the looping time $\tau$ is longer than the relaxation time of the chain, so all internal modes are fully relaxed before the two ends meet and SSS theory works well. 

Amitai {\it et al} \cite{Amitai2012-Flexible} calculate the perturbation of the spectrum for the Fokker-Planck equation \cref{F-P} and estimate the looping time accordingly. If we assume the natural boundary condition, the system is detailed balanced with the Boltzmann distribution as unique stationary density \cite{Qian-OU}. Also, in this case, the operator $L_F$ in the linear PDE \cref{F-P} is hermitian in the Hilbert space $L^2(\mX)$ with the inner product defined as $\langle f,g \rangle=\int_\Omega \frac{f(\mX)g(\mX)}{P_e(\mX)}\rd \mX$ \cite{VanKampen}. The linear operator can be rewritten as 
\begin{eqnarray}
L_FP(\mX,t)=\nabla\cdot \Big(P_e(\mX) \nabla \Big(\frac{P(\mX,t)}{P_e(\mX)}\Big)\Big)
\end{eqnarray}
Therefore it has a discrete real spectrum $\{-\lambda_i^0\}$ with complete orthogonal eigenfunction basis $\{\phi_i^0\}$. The spectrum consists of nonnegative integer-weighted sums of eigenvalues of the Laplace matrix, $-\gamma_p,$ and the eigenfunctions are products of the corresponding Hermite polynomials. The top eigenvalue $\lambda_0^0=0$ and the corresponding eigenfunction is the stationary density function. The next two are $\lambda_1^0=\gamma_1$ and $\lambda_2^0=\gamma_2$. 

But with the absorbing boundary condition, the spectrum $\{-\lambda_i^\epsilon\}$ is shifted due to removing the small tubular neighborhood $S_\epsilon$ and there is no longer a stationary density. If $\epsilon$ is small enough, the spectrum will be still distinct discrete and real. The time dependent solution can still be expressed as an expansion in the eigenfunctions $\{\phi_i^\epsilon\}$ with coefficients $c_i$,
$P(\mX,t)=\sum_{i=0}^\infty c_i \phi_i^\epsilon(\mX)\exp(-\lambda_i^\epsilon t)$. If the initial condition is the previous stationary density, i.e. the Boltzmann distribution, the eigenfunction $\phi_0^\epsilon$ is close to the Boltzmann distribution so that most of the energy is contributed by this zeroth mode, i.e, $c_0\approx 1$.  The survival probability $\Sigma(t)=\int_{\Omega-S_\epsilon} P(\mX,t)\rd \mX$, is in fact the summation of infinitely many exponential distributions. If the survival probability decays to 0 sufficiently fast, i.e, $\lim_{t\rightarrow +\infty} t\Sigma(t)=0,$ then the mean first passage time $\tau$ is simply the integral of the survival probability from 0 to $+\infty,$
 \begin{align} \label{AKH}
 \tau=\int_0^\infty\Sigma(t)\rd t =\sum_{i=0}^{\infty}\frac{C_i }{\lambda^\epsilon_i} \approx \frac{1}{\lambda^\epsilon_0}. 
\end{align}
Amitai {\it et al} \cite{Amitai2012-Flexible} discovered that the first passage time roughly follows one single exponential distribution for relatively small $N$ ($N<64$) by numerical simulation, so the zeroth mode in \Cref{AKH} is enough. We will discuss the distribution of the first passage time in  \Cref{numerical-simulation}. 
When $\epsilon$ is small enough, the first order perturbation of the zeroth eigenvalue is proportional to the ratio of the partition function of the closed polymer chain to the whole configuration space.  This ratio is again exactly the J factor and the proportionality spatial factor is $8\pi \epsilon$. 
\begin{align}\label{lambda-expansion}
\lambda_0^\epsilon=0+8\pi \epsilon J+O(\epsilon^2)
\end{align}
The series expansion of the eigenvalues follows results in \cite{ward2011}. The J factor in the scaling of $N$ appears in the expression was discussed in \cite{Guerin2013}.  If one uses \cref{lambda-expansion} in \cref{AKH}, then the first order perturbation result exactly matches the SSS Markovian approximation, i.e, the looping time is exactly \cref{SSS-approx}.

In summary, the SSS theory describes the kinetics of loop formation as a diffusion process in an effective potential of mean force that is derived from the equilibrium distribution for the end-to-end distance $P_e(r)$. It approximates the non-Markovian dynamics of the end-to-end distance $r$ by these simple Markovian dynamics.  
It assumes that the internal modes have relaxed before the looping process starts (this is called {\it the local equilibrium assumption}). The condition for this to hold is $\epsilon\ll 1/\sqrt{N} $. As far as the spectrum of the linear operator is concerned, this Markovian approximation corresponds to the first order perturbation of the zeroth mode. In \Cref{numerical-simulation}, it is shown that the SSS result significantly underestimates the looping time (it can also be proved through a variational principle \cite{Portman2003}). 
 Roughly speaking, the Markovian estimate ignores the case that two ends meet each other due to fluctuation, but the center of mass of the chain may be far away from the two ends. Intuitively, this case is more likely to happen for relatively large $\epsilon$ or large $N$. This drawback brings us to a more comprehensive method, Wilemski-Fixman theory. 
 
\section{WF theory}\label{WF}

One could incorporate a distance-dependent reaction term in the Fokker Planck equation \cref{F-P} with a microscopic rate constant $\kappa$ and relate the looping time with the time integral of a normalized sink-sink time correlation function. This is the celebrated Wilemski-Fixman (WF) approximation \cite{WilemskiFixman1, WilemskiFixman2}. Instead of solving a Fokker-Planck equation \cref{F-P} with a complicated boundary condition, the modified equation is of convection-diffusion-reaction type in free space (see \cref{F-P-mod}) and one can express the solution in terms of the Green function $G(\mX,t |\mX', 0)$ for \Cref{F-P}. This method can also be applied to study the Fokker-Planck equation with other difficult boundary conditions, like diffusion-limited catalytically-activated chemical reactions with a special catalytic subvolume \cite{Benichou2005}, facilitated diffusion and looping with a heterogenous  Rouse chain. 
The microscopic rate constant $\kappa$, which measures the effectiveness of the reaction, is not the same as the coefficient in the partially absorbing boundary condition in Collins and Kimball's kinetic theory. In fact, Szabo {\it et al}  \cite{Szabo1984} shows that the partially absorbing boundary condition is equivalent to the use of a delta sink on the reactive surface in conjunction with a reflecting boundary condition. However, as we will show later, if we let $\kappa\rightarrow +\infty$ in the WF approximation, the absorbing boundary condition can be recovered under some assumptions.  


\begin{equation}\label{F-P-mod}
\frac{\partial Q(\mX,t)}{\partial t}=L_FQ(\mX,t)-\kappa S(\mX)Q(\mX,t), \ Q(\mX, 0)=P_e(\mX)
\end{equation}
Two popular choices for the sink function are the Heaviside function and the radial delta function. For a given configuration $\mX$, the Heaviside function is
$S_1(\mX)= \begin{cases} 1, & \text{if}\ \|\mR_{ee}\|\le \epsilon\\ 0, & \text{Otherwise} \end{cases}$.
Intuitively, it corresponds to the Brownian particle starting to react once inside the reactive surface $S_\epsilon$. 
The delta function is 
$S_2(\mX)=\frac{1}{4\pi \epsilon^2}\delta(\|\mR_{ee}\|-\epsilon)$.
It is also called the Smoluchowski reaction sink. The reactive spherical surface is in three-dimensional space and it separates the whole space into two regions, inside and outside of the reactive surface. But this sink doesn't allow the Brownian particle to pass the surface. Both types of sinks have a three-dimensional representation in terms of the end-to-end vector $\mr$. 

The solution of \cref{F-P-mod} starting with the equilibrium distribution $P_e(\mX)$ can be expressed with the help of Dyson's formula as
\begin{align} \label{Q-solution}
Q(\mX, t)=P_e(\mX)-\kappa \int_0^t \rd t' \int \rd \mX' G(\mX, t|\mX', t' )S(\mX')Q(\mX', t' ). 
\end{align}
The inner space integral gives the probability density at time $t$ in the original Fokker-Planck equation \cref{F-P} if the initial distribution at time $t'$ is $S(\mX)Q(\mX, t')$. The meaning of \cref{Q-solution} is as follows: the probability of a chain to have the configuration $\mX$ at time $t$, is the probability to observe the configuration $\mX$ without sink, minus the probability of reaching the configuration $\mX$ at time $t$ starting in the sink at some point $t'$ between 0 to $t.$ The advantage of this approach is that the Green function for \cref{F-P} is known. The survival probability $\Sigma(t)$ is given by 
$\Sigma(t) =1- \kappa\int_0^t \rd t' \int \rd \mX' S(\mX')Q(\mX', t' )$.

To solve analytically for $Q(\mX, t)$ is still very difficult. The WF theory takes advantage of the conditional probability $T$ to reduce the dimensionality  from $\mathbb{R}^{3N+3}$ to $\mathbb{R}^3$. One can multiply $S(\mX)$ on both sides of \cref{Q-solution} and integrate w.r.t. $\mX$ to get \cref{Q-solution-2}. If one multiplies with a sink form other than $S(\mX),$ the result will be an unbalanced sink-sink correlation.
\begin{align} \label{Q-solution-2}
\int Q(\mX, t)S(\mX) \rd \mX=& \int P_e(\mX)S(\mX) \rd \mX - \kappa\int_0^t \rd t' \int \rd \mX S(\mX) \\ \nonumber
&\times\int \rd \mX'  G(\mX, t|\mX', t' )S(\mX')Q(\mX', t' ) 
\end{align}
If $S(\mX)Q(\mX, t)$ is the canonical ensemble with the Boltzmann distribution for the end-to-end vector $\mr,$ one could rewrite \cref{Q-solution-2} in terms of $T(\mr, t| \mr_0, 0).$ Specifically, for the Heaviside sink, the space and time dependencies of $Q(\mX, t)$ are separated inside the sink, $Q(\mX, t)=P_e(\mX)g(t,\kappa)$ when $\|\mr\|\le \epsilon$; for the delta sink, $Q(\mX, t)$ only requires the same space and time separation on the reactive surface, i.e, $Q(\mX, t)=P_e(\mX)g(t,\kappa)$ when $\|\mr\|=\epsilon$. The condition for the delta sink is weaker than for the Heaviside sink, so one would expect that the delta sink performs better than the Heaviside sink. Note that both are weaker than the original WF  assumption which requires time and space separation in the whole space. In either case, $Q(\mX, t)$ is not homogenous on the reactive surface. 
 In fact, WF theory allows non-Markovian dynamics for the end-to-end vector of the chain but doesn't capture the full non-Markovian effect.
  
With the help of the sink-sink correlation function $C(t)$, \Cref{Q-solution-2} can be rewritten into a more compact form \cref{sink-sink} and \cref{Q-solution-3}, if the separation condition is satisfied. 
\begin{align} \label{sink-sink}
&C(t)=\langle S(\mr,0),S(\mr,t)\rangle=\int \rd \mr \int \rd \mr' S(\mr) T(\mr, t| \mr', 0)S(\mr')P_\mr(\mr')  \\ \label{Q-solution-3}
&g(t,\kappa)P_0 = P_0 - \kappa \int^t_0 g(t',\kappa)C(t-t')\rd t', \ \ P_0=\int P_\mr(\mr)S(\mr) \rd \mr.
\end{align}
One observes that $C(\infty)=(P_0)^2$ for both sinks. For the delta sink, $C(t)$ has a singularity at $t=0$; for the Heaviside sink, $C(0)=P_0$, the probability of the looped state at equilibrium. In general, there is no closed form expression for $C(t)$. 

Assume that on the reactive surface, the time function $g(t, \kappa)$ is asymptotically  $q(t)/\kappa+O(1/\kappa^2)$. If we let the microscopic rate constant $\kappa\rightarrow +\infty$, then $Q(\mX, t)\rightarrow 0$ on the reactive surface asymptotically and the absorbing boundary condition is recovered.  Under this assumption, the survival probability will be 
$\Sigma(t)=1-\Big(\int_0^t P_0q(t')\rd t'\Big).$ In this limit, the left hand side of \cref{Q-solution-3} vanishes and we find that
\begin{align} \label{deconvolution}
P_0 =\int^t_0 q(t')C(t-t')\rd t'.
\end{align} 
\Cref{deconvolution} is valid inside the sink for the Heaviside sink and on the reactive surface for the delta sink respectively, and looping time only requires the solution in these regions. This defines a deconvolution problem with the kernel given by $C(t).$ This is an inverse problem and one way to solve it is to use the Laplace transform. The Laplace transform of $q(t)$ and $\Sigma(t)$ are, $\hat{q}(s)=\frac{P_0}{s\hat{C}(s)}$ and $\hat{\Sigma}(s)=\frac{1}{s}-\frac{(P_0)^2}{s^2\hat{C}(s)}$ resepctively.
The looping time is given by $\tau=\lim_{s\rightarrow 0}\hat{\Sigma}(s).$ Since $C(t)$ is a decreasing function to $C(\infty)$, in the transform variable we have $\hat{C}(s)>C(\infty)/s.$ If one approximates $\hat{\Sigma}(s)$ in the denominator by $sC(\infty),$ then $\tau\le\lim_{s\rightarrow 0}\frac{\hat{C}(s)}{C(\infty)}-\frac{1}{s}.$ In fact, we can show the inequality is an equality as follows. 

Define $I(t)=\frac{C(t)}{C(\infty)}-1.$ The improper integral of $I(t)$ from 0 to $+\infty$ is finite, and it is equivalent with $\hat{I}(0)< \infty$ in Laplace transform. The looping time $\tau$ is given by
\begin{align} \nonumber
\tau&=\lim_{s\rightarrow 0} \frac{s\hat{C}(s)-(P_0)^2}{s^2\hat{C}(s)}= \lim_{s\rightarrow 0}\frac{\hat{I}(s)}{s\hat{I}(s)+1} \\ \label{WF-appro}
&=\hat{I}(0)=\int_0^\infty \rd t\Big(\frac{C(t)}{C(\infty)}-1\Big)=\lim_{s\rightarrow 0}\frac{\hat{C}(s)}{C(\infty)}-\frac{1}{s}.
\end{align}
Finally, the integral $\int_0^\infty \rd t\Big(\frac{C(t)}{C(\infty)}-1\Big)$ is the famous WF approximation formula which one can evaluate numerically. But the integrand $\frac{C(t)}{C(\infty)}-1$ is clearly not the survival probability $\Sigma(t)$ because this small alteration in the Laplace domain would change the integrand function significantly in the time domain.

We can provide a probabilistic interpretation of the WF formula for the Heaviside sink. $C(0)=P_0$ is   the fraction of the polymer that has already formed the loop initially. If one only takes this fraction of the polymers to start the process, the polymer will start to un-loop gradually. We assume that once they are un-looped, these polymers will not form the loop again under the time scale of interest. After sufficiently long time, the distribution of end-to-end vector will be the Gaussian distribution with normalization factor $P_0$ and the fraction of polymers that is still in the loop state will be $C(\infty)=P_0^2.$ So C(t) describes the fraction of the polymer that is still in the loop state at time $t$ if the process starts with the looped polymer. The process of the polymers un-looping themselves is intimately related to the looping problem. We can rewrite the integrand as follows
 \begin{align}
 \tau= \frac{1-C(0)}{C(0)}\int_0^\infty \frac{C(t)-C(\infty)}{C(0)-C(\infty)} \rd t.
 \end{align}
The expression $\frac{C(t)-C(\infty)}{C(0)-C(\infty)}$ is taken as the approximation of the survival probability for the un-looping process. The integral calculates the expected un-looping time. If one considers this looping process as the dynamical process in a bi-stable system, the ratio $\frac{1-C(0)}{C(0)}$ is the relative stability of the two potential wells, which builds the connection from un-looping time to looping time. A disadvantage with this argument is that it does not work for the delta sink case. 

In summary, the approximations made are: 

(i) The solution $Q(\mX, t)$ of \cref{F-P-mod} has the space and time separation inside the sink for the Heaviside sink and on the reactive surface for the delta sink. This approximation is the key step for the construction of this conditional probability w.r.t. canonical ensemble and the reduction in dimension. It is a plausible assumption because the canonical ensemble of polymers will behave as quasi-stationary, at least for the small capture radius case. In order to go beyond the WF approximation, one has to relax this condition. 

(ii) The time function $g(t,\kappa)$ is analytic with respect to the variable $1/\kappa$, such that the absorbing boundary is recovered. Although the assumption is difficult to verify, it is reasonable physically. As the microscopic rate constant $\kappa$ increases, the reaction is more and more likely to happen once the two ends are within the capture radius. In the limit, the reaction will happen immediately which corresponds to the absorbing condition. 

(iii) Instead of solving the deconvolution problem numerically, the WF theory approximates the denominator in the Laplace domain, and it gives a semi-analytical form for the looping time $\tau$ directly, but the analytical form for the survival probability is unknown. We will address this issue in \Cref{numerical-simulation} and solve numerically for the survival probability $\Sigma(t)$ there.  

\section{Perturbation theory}\label{perturb-section}
Although the WF theory provides a good estimate of the looping kinetics, the numerical integration neither provides a reduced model nor gives the scaling law in the two parameters $N$ and $\epsilon.$ We are going to apply perturbation techniques on the time integral to extract asymptotic estimates of the looping time. 

The sink-sink correlation function $C(t)$ can be expressed as a double integral over two radial variables r and $r',$ after averaging out the azimuthal and polar angles. It uses the fact that $\int \rd \mr \int \rd \mr_0 \exp(\mr\mr')=\int \rd r 4\pi r^2 \int \rd r' 4\pi r'^2 \frac{\sinh(rr')}{rr'}$.
\begin{align}
C(t)=&\frac{(3/2\pi N )^3}{(1-\phi^2)^{3/2}}\int_0^\infty \rd r 4\pi r^2S(r) \int_0^\infty \rd r' 4\pi r'^2S(r') \exp\Big(-\frac{3}{2N }\frac{r^2+r'^2}{1-\phi^2}\Big) 
\\ \nonumber
&\times \sinh\Big(\frac{3\phi rr'}{N (1-\phi^2)}\Big)/\frac{3\phi rr'}{N (1-\phi^2)}
\end{align}
We introduce the small dimensionless quantity $x_0=\frac{3\epsilon^2}{2N }\ll 1.$ $x_0$ is a small quantity since the nature of the looping assumes the capture radius $\epsilon$ is much smaller than the average end-to-end distance $\sqrt{N} $. Using $x_0$ we can find explicit expressions for the integrand $I(t).$

For the Heaviside sink, the double integral is evaluated by expanding in powers of $x_0$, 
$C(t)=\frac{16x_0^3}{9\pi(1-\phi^2)^{3/2}}\Big( 1-\frac{6x_0}{5(1-\phi^2)}+\dots\Big)$.
To match the similar form in \cite{Doi1975}, Pastor {\it et al} \cite{Szabo1996} proposed the closed form
$C(t)\approx \frac{16x_0^3}{9\pi}\Big(1-\phi^2+\frac{4}{5}x_0\Big)^{-3/2}$,
which matches the first two order of the expansion in $x_0$. Then 
\begin{align}
\label{Ct_heavi}
I_{H}(t)=\frac{C(t)}{C(\infty)}-1\approx  \Big(\frac{1+\frac{4}{5}x_0}{1-\phi^2(t)+\frac{4}{5}x_0}\Big)^{3/2}-1
\end{align} 
The time integral of $I_H(t)$ is roughly approximated in two different time scales. First, in the short time scale, $\phi\approx 1$, 
the denominator is approximated by $ 2(1-\phi)+\frac{4}{5}x_0.$ The integrand in this time scale is much larger than 1 since $x_0$ is sufficiently small. 
The approximation of $I_H(t)$ is, $I_H(t) \approx  \Big(\frac{1+\frac{4}{5}x_0}{2(1-\phi(t))+\frac{4}{5}x_0}\Big)^{3/2}$. Second, in the long time scale, $\phi\approx 0$, one could estimate the asymptote as the integrand goes to 0 by exploiting that the quantity  $\frac{\phi^2(t)}{1-\phi^2(t)+\frac{4}{5}x_0} $ is small. So the approximation of $I_H(t)$ under this time scale is, $I_H(t)\approx \frac{3}{2}\frac{\phi^2(t)}{1+\frac{4}{5}x_0}$.

For the delta sink, the double integral is evaluated exactly
\begin{align} \nonumber
&C(t)=\frac{12x_0/N\pi}{\phi\sqrt{1-\phi^2}} \exp\Big(\frac{-2x_0}{1-\phi^2}\Big)\sinh\Big(\frac{2x_0\phi}{1-\phi^2}\Big), C(\infty)=\frac{24x_0^2}{N \pi} \exp(-2x_0) \\ \label{Ct_ds}
&I_{DS}(t)=\frac{C(t)}{C(\infty)}-1=\frac{\exp\Big(2x_0\phi/(1+\phi)\Big)-\exp\Big(-2x_0\phi/(1-\phi)\Big)}{4x_0\phi\sqrt{1-\phi^2}} -1.
\end{align}
Similarly to the Heaviside sink case, it can be approximated in two different time scales. First, in the short time scale, $\phi\approx 1$, 
 so the approximation of $I_{DS}(t)$ is given by $I_{DS}(t)\approx \frac{\exp(x_0)-\exp(-2x_0/(1-\phi))}{4\sqrt{2}x_0\sqrt{1-\phi}}.$ Second, in the long time scale, $\phi\approx 0,$ one can use a Taylor expansion to estimate the limit of the integrand. So the approximation of $I_{DS}(t)$ is $I_{DS}(t)\approx \frac{3}{2}\phi^2$. 

Since the time correlation function for the end-to-end vector $\phi(t)$ doesn't have a closed analytical form, the integral is still not feasible analytically. We will approximate $\phi(t)$ in three different time scales: First, when the time scale is within the relaxation time for the largest mode, $t\le \tau_N=\frac{1}{12}$, take the approximation $\exp(-\gamma_p t)=1-\gamma_p t+ O(t^2)$. Then $\phi(t)$ can be approximated by $\phi(t)\approx 1-\frac{6 t}{N }$. Second, when time $t$ between two time scales, $\tau_N\ll t \ll \tau_3=\frac{N^2 }{27\pi^2}$, the approximation will be based on $\frac{1}{\tan^2(\theta_p)}\approx \frac{1}{\theta_p^2}$ and $\gamma_p\approx p^2\gamma_1$. Both approximations work well on small $p$ that also contribute the most to the correlation function, $\phi(t)\approx 1-\frac{8}{\pi^2}\sum_{p\ \text{odd}}\frac{1}{p^2}\Big(1-\exp(-p^2\gamma_1 t)\Big)$.
One has to further approximate by turning the summation into the integral. Define $x=p\sqrt{\gamma_1  t}$, the coefficient $N^{-1}\ll \sqrt{\gamma_1 t} \ll 1$ is a small quantity, 
$\phi(t)\approx 1-\frac{4}{\pi^2}\sqrt{\gamma_1  t}\int_0^{+\infty}\frac{1}{x^2}(1-\exp(-x^2))\rd x=1-\frac{4}{N}\sqrt{\frac{3 t}{\pi }}$.
Last, when the time $t\gg \tau_3,$ all other modes are relaxed except the first mode. Then it can be approximated by $\phi(t)\approx \frac{8}{\pi^2} \exp(- \gamma_1 t)$. 
In summary, the end-to-end vector correlation function $\phi(t)$ has the following analytical approximation 
\begin{equation}
\phi(t) \approx \begin{cases}
  1-\frac{6t}{N } & \text{Short Timescale} \\
 1-\frac{4}{N}\sqrt{\frac{3t}{\pi }} &\text{Median Timescale} \\
  \frac{8}{\pi^2}\exp(-\gamma_1 t) & \text{Long Timescale} 
\end{cases}
\end{equation}
The effective diffusion coefficient $D_{eff}$ for the end-to-end vector is defined by the end-to-end correlation function $\phi(t)$, 
\begin{equation}
D_{eff}(t)=\frac{\langle (\mR_{ee}(t)-\mR_{ee}(0))^2\rangle}{6t}=\frac{N (2-2\phi(t))}{6t}
\end{equation}
In the short timescale, $D_{eff}=2$ which is time homogeneous. It also verifies the choice of diffusion coefficient in SSS theory. SSS theory doesn't capture the behavior from other timescales, so it cannot reproduce the mixed scaling law.

Numerical simulations show that all three asymptotic results perform very well under the appropriate timescale. Specifically, for $N=75,$ the short timescale approximation works better than the median for $t\le t_1=\frac{4}{3\pi}\approx 5.1\tau_N $ and the long timescale approximation is better than the median for $t\ge 2.8\tau_3$. Notice when $t\approx \tau_3,$ the correlation function $\phi$ is about 0.76, so one can still treat $\phi\approx 1$ under the median timescale. Then $I_H$ and $I_{DS}$ have analytical approximations $\bar{I}_{H}$ and $\bar{I}_{DS}$ as follows,
\begin{equation}\label{perturbation}
\bar{I}_H(t)= \begin{cases} 
\Big(\frac{1+\frac{4}{5}x_0}{\frac{12t}{N }+\frac{4}{5}x_0}\Big)^{3/2} & \text{Short Timescale} \\
\Big(\frac{1+\frac{4}{5}x_0}{\frac{8}{N}\sqrt{\frac{3t}{\pi }}+\frac{4}{5}x_0}\Big)^{3/2} & \text{Median Timescale} \\ 
\frac{96}{\pi^4(1+\frac{4}{5}x_0)}\exp(-\frac{6\pi^2t}{(N+1)^2}) &\text{Long Timescale} 
\end{cases} 
\end{equation}  

\begin{equation}\bar{I}_{DS}(t)=\begin{cases} 
\frac{\exp(x_0)-\exp(-\frac{N x_0}{3 t})}{8\sqrt{3}x_0 \sqrt{\frac{t}{N }}} & \text{Short Timescale}  \\
\frac{\exp(x_0)-\exp(-\frac{N x_0}{2}\sqrt{\frac{\pi }{3 t}})}{8\sqrt{2}x_0\sqrt{\frac{1}{N}\sqrt{\frac{3t}{\pi }}}} &\text{Median Timescale} \\
\frac{96}{\pi^4}\exp(-\frac{6\pi^2t}{(N+1)^2}) & \text{Long Timescale} 
\end{cases}
\end{equation}

\begin{figure}
\begin{tabular}{cc}
{\bf (a)}  \includegraphics[valign=t, scale=0.3]{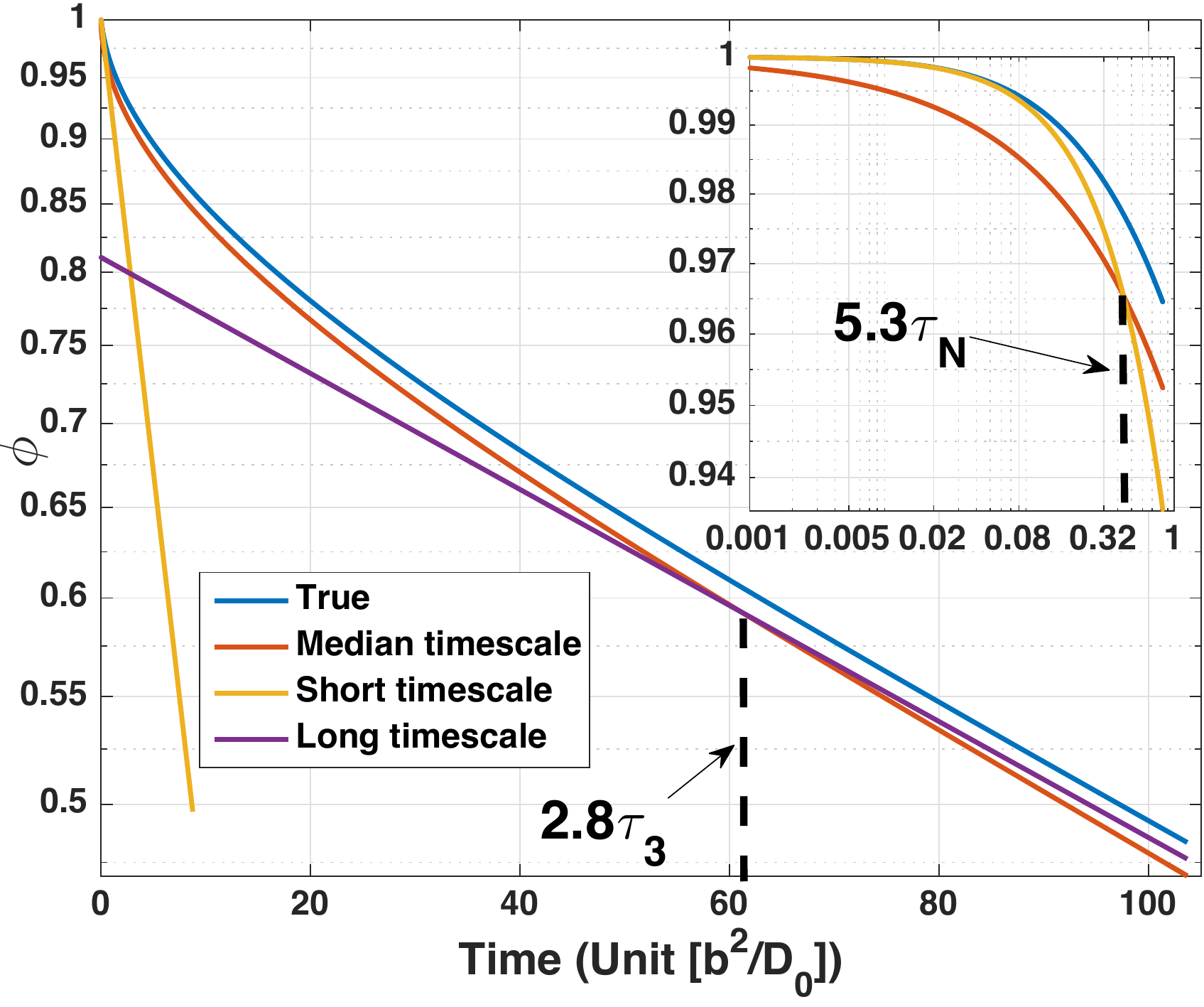} &{\bf (b)} \includegraphics[valign=t, scale=0.3]{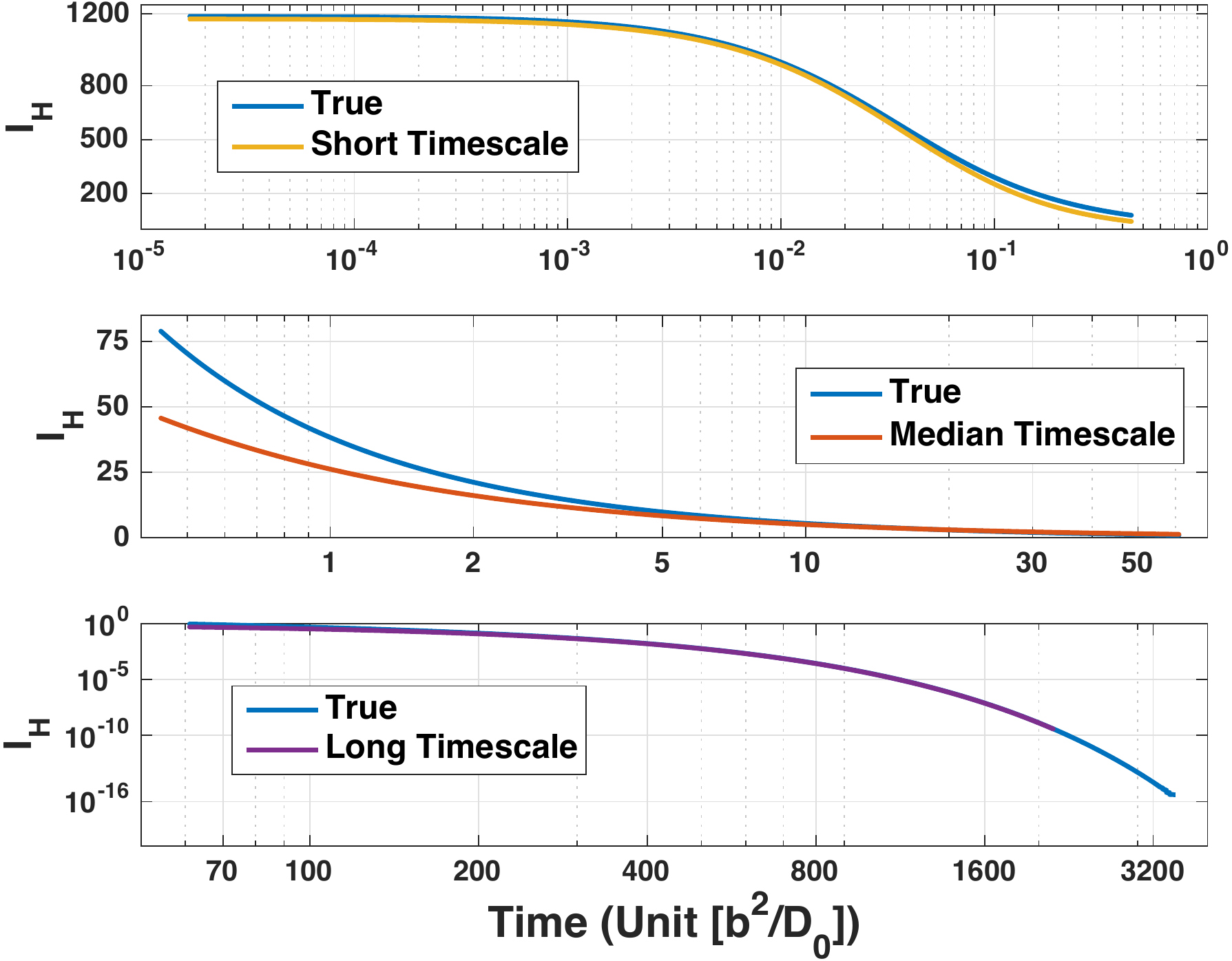}\\
\end{tabular}
\begin{tabular}{c}
\ \ \ \ \ \ \ \ \ \ \ \ \ \ \ \ \ \ \ \ \ \ \ \ \ \ \ \ {\bf (c)} \includegraphics[valign=t,scale=0.3]{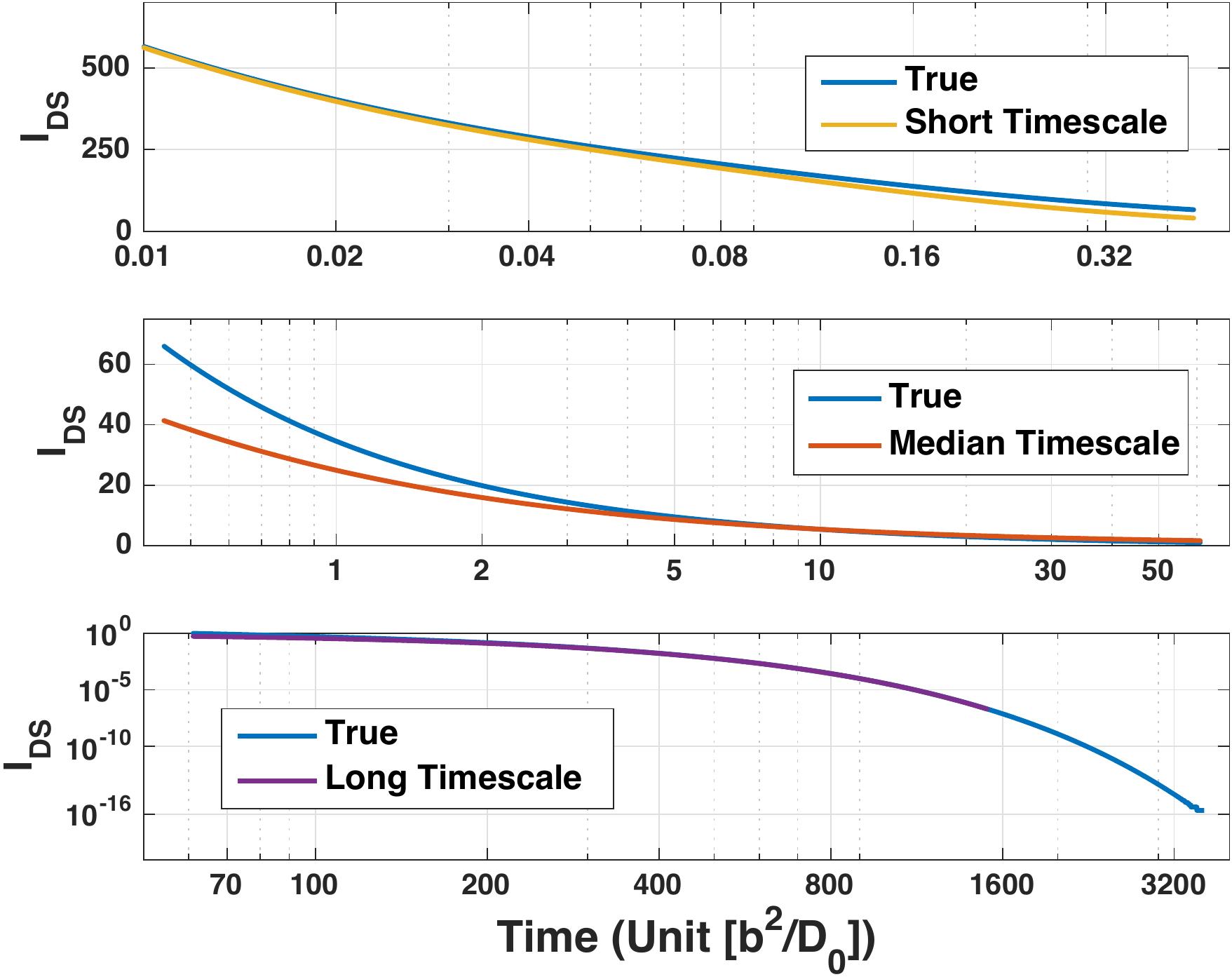}
\end{tabular}
\caption{(a) Comparison of numerical calculation and analytical approximations for the end-to-end vector correlation function $\phi(t)$ under three different timescales for $N=75$. (b)  The approximated integrand $\bar{I}_H(t)$ from \cref{perturbation} and compared to the numerical evaluation from \cref{Ct_heavi} for $N=75$ and $\epsilon=0.75$. (c)  The approximated integrand $\bar{I}_{DS}(t)$ from \cref{perturbation} and compared to the numerical evaluation from \cref{Ct_ds} for $N=75$ and $\epsilon=0.75$. }
\end{figure}

Specifically, for $N=75$ and $\epsilon=0.75$, the analytical approximations perform very well for both sinks. The time range within which they deviate most is on the boundary of the short timescale and the median timescale. In terms of numerical integration, most of deviation is contributed by the median timescale and its boundaries as expected (figure not shown). 
The advantage of the approximation is that the integral which provides an estimate of the looping time can be evaluated analytically. Set the two timescale break points as $t_1= \frac{4}{3\pi }$ and $t_2=3\tau_3$. The second breakpoint between the median and long timescales depends on the number of beads, $N.$ We roughly choose $3\tau_3$ as the second breakpoint but this will not change the behavior of scaling law qualitatively. 

For the Heaviside sink, one can compute the mixed scaling law explicitly,
\begin{align} \label{heavi_anal}
&\int_0^\infty\bar{I}_H(t)\rd t=\frac{N\sqrt{N}}{\epsilon}\Big(\frac{1}{6}\sqrt{\frac{5}{6}}(1+\frac{4}{5}x_0)^{3/2}\Big)\Big(1-\sqrt{\frac{6\epsilon^2/5}{16/\pi+6\epsilon^2/5}}\Big)\\ \nonumber
&+(N+1)^2\frac{16}{\pi^6(1+\frac{4}{5}x_0)}\exp(-\frac{2}{3}\frac{N^2}{(N+1)^2})+N\sqrt{N}\Big(\frac{\pi}{48}(1+\frac{4}{5}x_0)^{3/2}\Big)\\  \nonumber
&\times\Big(\sqrt{\frac{8N}{\pi}\sqrt{\frac{1}{3\pi}}+\frac{6\epsilon^2}{5}}-\sqrt{16/\pi+6\epsilon^2/5}+\frac{6\epsilon^2/5}{\sqrt{\frac{8N}{\pi}\sqrt{\frac{1}{3\pi}}+\frac{6\epsilon^2}{5}}}-\frac{6\epsilon^2/5}{\sqrt{16/\pi+6\epsilon^2/5}}\Big)
\end{align}
When the capture radius $\epsilon\ll 1,$ this integral is 
$\int_0^\infty\bar{I}_H(t)\rd t\approx h_1\frac{N\sqrt{N}}{\epsilon}+h_2N\sqrt{N}+h_3N^2$
where $h_1,h_2$ and $h_3$ are constants. In the limit of $N\rightarrow +\infty$, we can ignore the middle term $h_2N\sqrt{N}$ and recover the mixed scaling law hypothesized before.  

For delta sink, the integral is much more complicated, 
\begin{align}\nonumber
I_1=&\frac{N\sqrt{N}}{\epsilon^2}\frac{1}{9\sqrt{\pi}}\Big(\exp(x_0)-\exp\Big(-\frac{3\pi \epsilon^2}{8}\Big)\Big) +\frac{N\sqrt{N}}{\epsilon} \frac{\sqrt{\pi}}{6\sqrt{6}} \Big(1-\erf\Big(\sqrt{\frac{3\pi\epsilon^2}{8}}\Big)\Big) \\ \nonumber
I_2=&\frac{N^3}{\epsilon^2}\frac{1}{ 2^{1/2} 3^{15/4} \pi ^{5/4}}\Big(\exp(x_0)-\exp\Big(-\frac{\sqrt{3\pi^3}}{2} x_0\Big)\Big)-N\sqrt{N}  \frac{\sqrt{\pi }}{18} \exp\Big(-\frac{3\pi \epsilon^2}{8}\Big) \\  \nonumber
&+N^2 \frac{\sqrt[4]{\frac{\pi }{3}}}{18 \sqrt{2}} \exp\Big(-\frac{\sqrt{3\pi^3}}{2} x_0\Big)-\frac{N\sqrt{N}}{\epsilon^2}\frac{2}{27 \sqrt{\pi }} \Big(\exp(x_0)-\exp\Big(-\frac{3\pi \epsilon^2}{8}\Big) \Big)\\ \nonumber
&+ N\sqrt{N}\epsilon \frac{\pi^{3/2}}{12\sqrt{6}}\Big(\erf\Big(\frac{\epsilon}{\sqrt{N }}\frac{(3\pi)^{3/4}}{2}\Big)-\erf\Big(\frac{(3\pi)^{1/2}}{2\sqrt{2}}\epsilon\Big)\Big).  \\ \nonumber
I_3=&(N+1)^2\frac{16}{\pi^6}\exp\Big(-\frac{2}{3}\frac{N^2}{(N+1)^2}\Big)
\end{align}
\begin{align}  \label{DS_anal}
&\int_0^\infty\bar{I}_{DS}(t)\rd t=I_1+I_2+I_3
\end{align}
Similarly when $\epsilon \ll 1$, one can use Taylor expansions for $\exp(x)$ and $\erf(x)$ functions at $x=0$. The integral is roughly 
$\int_0^\infty\bar{I}_{DS}(t)\rd t \approx  d_1 \frac{N\sqrt{N}}{\epsilon}+d_2 N\sqrt{N}+d_3N^2$ as well,
where $d_1, d_2$ and $d_3$ are constants. In the limit $N\rightarrow +\infty$, it recovers the mixed scaling law again. Specifically, $d_1$ is mostly contributed by the short timescale approximation in $I_1$ and is about $\frac{\sqrt{\pi}}{6\sqrt{6}}$. It exactly matches the SSS result as predicted in \cite{Szabo1996}. It implies that the SSS and WF theories give the same asymptotic result $\frac{N\sqrt{N}}{\epsilon}$ in the limit $\epsilon\rightarrow 0$ for given $N$. In practice, this asymptotic result is realized when $\epsilon$ is extremely small. 
$d_2$ and $d_3$ are mostly contributed by the median and long timescale approximations. This is also  predicted by Doi in \cite{Doi1975} and Doi provided a dynamical explanation as well. So both $ N\sqrt{N}$ and $N^2$ term are considered as the next order approximation result when $\epsilon$ is still relatively small. One would insightfully rewrite the looping time with the mixed scaling law as $\tau=\frac{N\sqrt{N}}{\epsilon}\Big(d_1+(d_2+d_3\sqrt{N})\epsilon +O(\epsilon^2)\Big)$. Note in the derivation of this scaling law, we assume $N$ as a constant is much larger than 1, i.e, $N\gg1$.  From the scaling law, two seemingly contradicting results from Doi and SSS are the consequence of the non-uniforming convergence of $\epsilon\to 0$ and $N\to +\infty$. Such results are typical in singular perturbation problems.

\section{Numerical Simulation}\label{numerical-simulation}
 
 \begin{figure}
\includegraphics[scale=0.35]{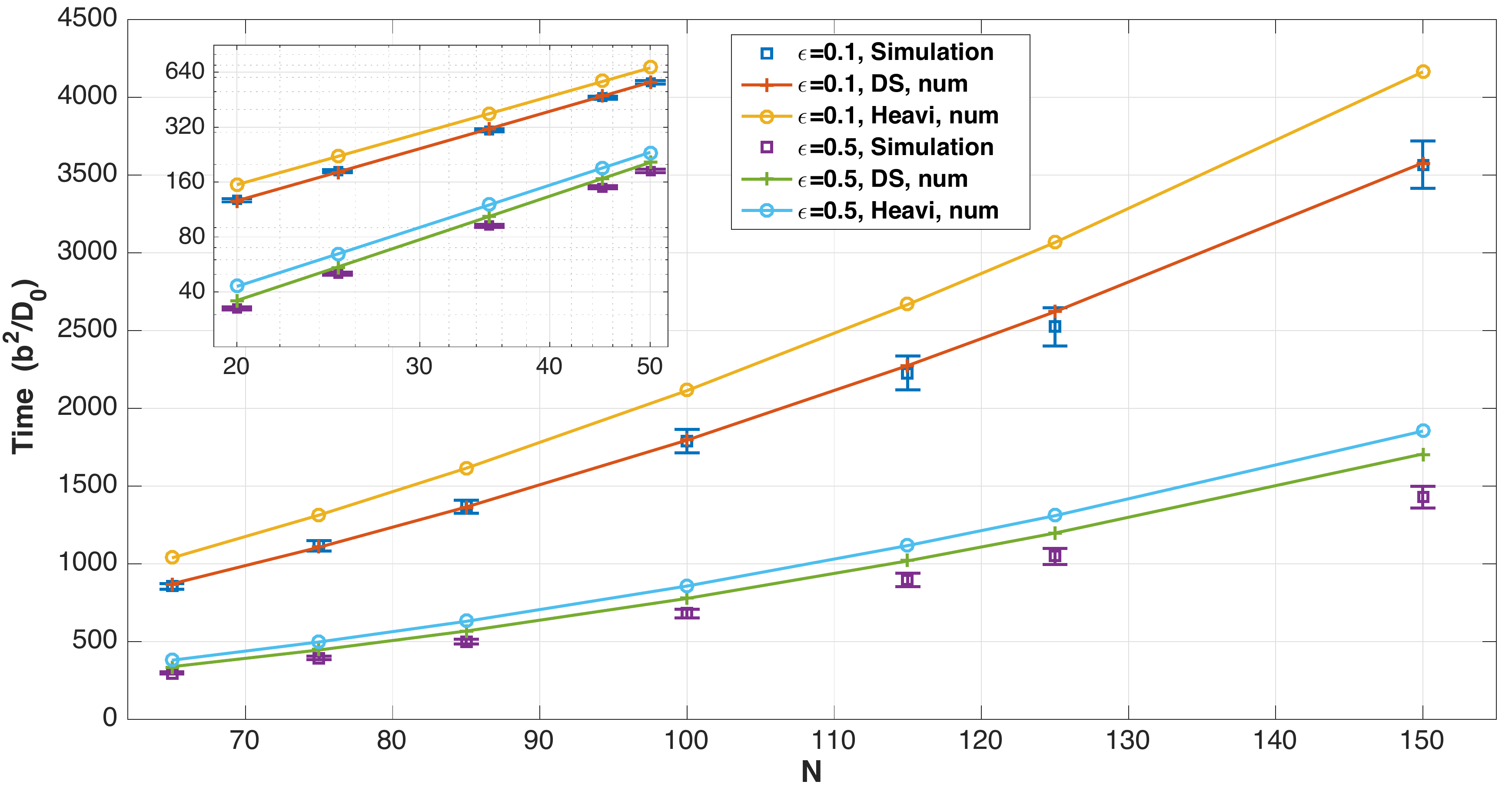}
\caption{Dependence of looping time on $N$ for two capture radius of $\epsilon=0.1$ and $\epsilon=0.5$. The looping time is estimated from a Monte Carlo simulation and compared to the numerical integration results in WF theory using \cref{Ct_heavi} and \cref{Ct_ds}. }\label{comparison}
\end{figure}

\begin{figure} 
\begin{tabular}{cc}
 \includegraphics[valign=t, scale=0.31]{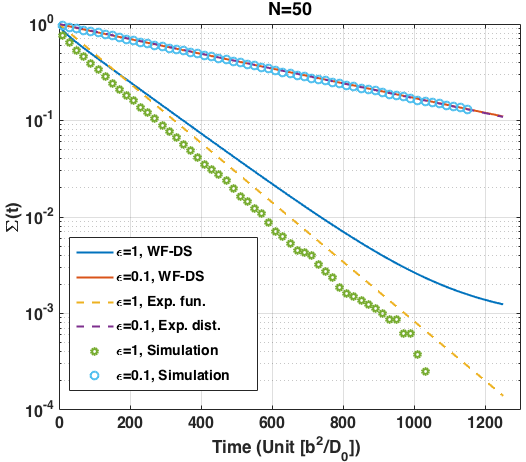} &\includegraphics[valign=t, scale=0.31]{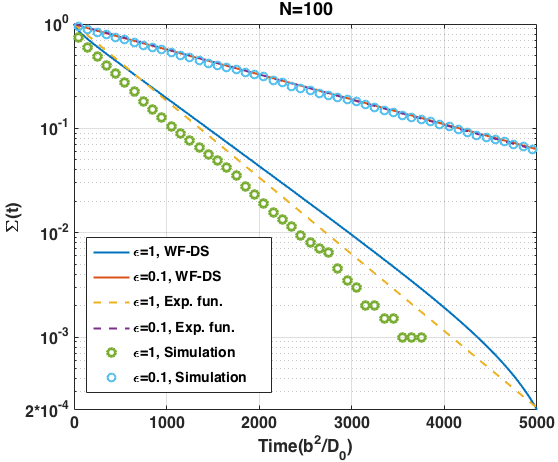}\\
\end{tabular}
\caption{Comparison of survival probability from WF theory with delta sink and from simulations for different $N$ and different $\epsilon$. The exponents of the exponential functions are the inverses of the looping time from WF theory in delta sink according to \cref{Ct_ds}.}
\label{survival}
\end{figure}

The Monte Carlo simulation algorithm is based on \cite{Szabo1996} but with more sample points and smaller time step. The simulation results are also within the range of their paper. The numerical solver is the Euler-Maruyama method, and the initial condition is sampled from the equilibrium distribution. The trajectory is terminated once the end-to-end distance is within the capture radius and record the passage time. The time step is chosen adaptively: when the end-to-end distance is within 2 Kuhn lengths, much finer time step is picked to prevent overshooting; but a larger time step is permitted when the end-to-end distance is outside of the range. The time step defined is 
\begin{equation}
\Delta t=\begin{cases}
\Delta_l+\Delta_h\sin\Big(\frac{\pi}{2}\Big(\frac{\|\mr\|-\epsilon}{2-\epsilon}\Big)^2\Big) & \epsilon\le\|\mr\|\le 2\   \\
\Delta_l+\Delta_h & \|\mr\|>2 \
\end{cases}
\end{equation}
with $\Delta_l=10^{-6}$ and $\Delta_h=10^{-3}$. This choice of time step is finer than that in \cite{Szabo1996}. For each parameter set, there are at least $n=2000$ samples. The passage time is considered to follow a single exponential distribution in \cite{Szabo1996}, so the 95\% confidence interval for the mean passage time is $ \bar\tau\pm\frac{1.96}{\sqrt{n}}\bar\tau$. We include this confidence interval for comparison. In fact, one can use the Lilliefors test, which is an improvement of the Kolmogorov-Smirnov test, to return a test decision for the null hypothesis that the passage time comes from an exponential distribution. The result is 1 if the test rejects the null hypothesis at 1\% significant level, 0 otherwise \cite{conover1999practical}. We applied the test to our simulation data for various parameters (we used $N=20,25,35,45,50,65,75,85,100,115,125,150$ and $\epsilon=0.1,0.25,0.5,0.75,1,2$ for a total of 72 parameter combinations). The test shows: when $\epsilon=0.1$ for all $N$ and $\epsilon=0.25$ with $N\ge 100,$ the test does not reject the null hypothesis at 1\% significant level; for other parameter ranges, the test rejects the null hypothesis and that means the data do not fit an exponential distribution. Our result is different from Amitai's result for large $N,$ where they claim that the passage time does not follow a single exponential distribution when $N>64.$ In fact, we can visualize the survival probability for the large $N$ case in \Cref{survival}. 
\begin{table}
 \begin{center}
      \begin{tabular}{|c|c|c |c |c |c |c|c|}
        \hline
$N$, $\epsilon$&Sample points&Simulation\ & H.n. &H.a.&DS. n. & DS. a. &SSS\\ \hline 
50, 0.1 &8000&563$\pm$12& 677 & 632 & 564 & 530&426 \\ 
50, 0.5 &8000& 184$\pm$4&233 &200& 205 & 186&85 \\ 
50, 1.0 &8000& 113$\pm$2 & 163 & 143 & 141 & 134&43 \\ \hline
75, 0.1 &4000& 1117$\pm$34 &1314& 1230&1107 & 1048 &783 \\ 
75, 0.5 &4000&396$\pm$12&498&437&447&414&157 \\ 
75, 1.0 &4000&261$\pm$8&368 &330 &326 & 317&783 \\ \hline
100, 0.1 &2000&1790$\pm$76&2114 &1984& 1796 &1710 &1206\\ 
100, 0.5 &2000&681$\pm$29&858 & 762 &778 & 733&241 \\
100, 1.0 &2000&466$\pm$20& 655 &595&590& 581&121 \\ \hline
\end{tabular}
\end{center}
\caption{Comparison of theoretical results and simulations for selected values of $N$ and the capture radius $\epsilon$. H. n. (Heaviside numerical) and DS. n. (Delta sink numerical) are obtained from the numerical integration in WF theory using \cref{Ct_heavi} and \cref{Ct_ds}. H. a. (Heaviside analytical) and DS. a. (Delta sink analytical) are analytical results from \cref{heavi_anal} and \cref{DS_anal}. SSS is the analytical result from \cref{SSS-approx}.} \label{table}
\end{table}

 \Cref{table} and \Cref{comparison} show that the WF theory overestimates the mean passage time, as can also be proved by using a variational principle \cite{Portman2003}. They also verify our argument that the delta sink should perform better than the Heaviside sink. 
 The analytical results underestimate the numerical integration at various points in the parameter range, by less than 15 percent for Heaviside sink and by less than 10 percent for delta sink. 
From \Cref{table} and \Cref{comparison}, we see that the delta sink results are in remarkable agreement with the simulation for small capture radius $\epsilon$. This implies that the space and time separation approximation relies mostly on the small capture radius assumption and not on the large number of beads. So, it is reasonable to use the WF theory as the simulation results when $\epsilon \ll 1$. At the same time, both analytical results predict that the WF theory has the mixed scaling law with $N\sqrt{N}$, $\frac{N\sqrt{N}}{\epsilon}$ and $N^2$. We use the WF theory for both sinks under the parameter range $0.1\le \epsilon\le 0.15$ and $100 \le N\le 150$ to fit the scaling law. One could use simulation results to fit but the computational cost is enormous. The coefficient $d_1$ is estimated as 0.1225, very close to the analytically predicted result $\frac{\sqrt{\pi}}{6\sqrt{6}}\approx 0.1206$. 
 \begin{align} \label{Heavi-fit}
 &\int_0^\infty I_H(t) \rd t= 0.1536\ \frac{N\sqrt{N}}{\epsilon}-0.0982\ N\sqrt{N}+0.0677\ N^2+\epsilon_1\\ \label{DS-fit}
& \int_0^\infty I_{DS}(t) \rd t = 0.1225\ \frac{N\sqrt{N}}{\epsilon}-0.1060\ N\sqrt{N}+0.0677\ N^2+\epsilon_2
 \end{align}
 \begin{figure}
\begin{tabular}{cc}
{\bf (a)}  \includegraphics[valign=t, scale=0.29]{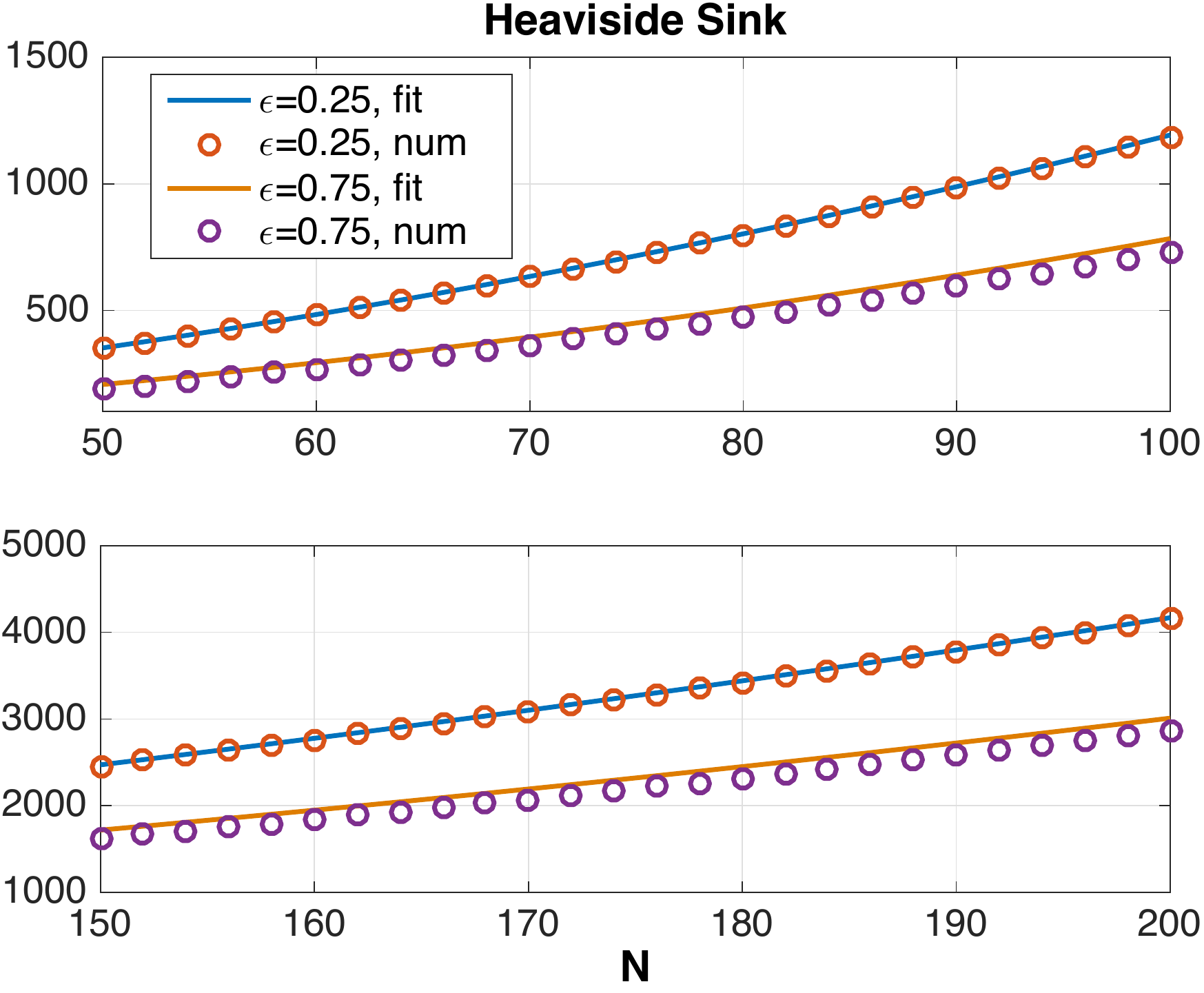} &{\bf (b)} \includegraphics[valign=t, scale=0.29]{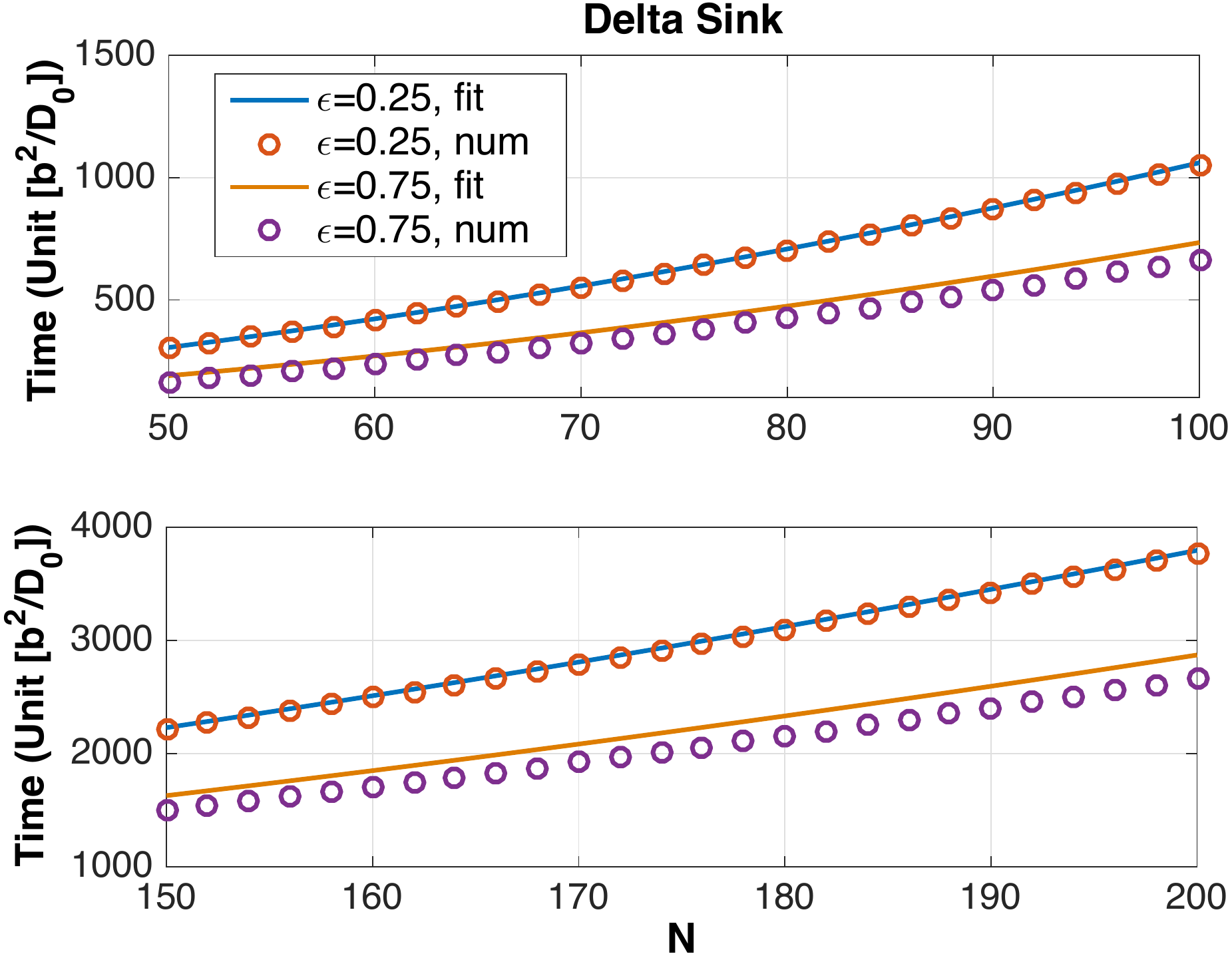}\\ 
{\bf (c)} \includegraphics[valign=t, scale=0.29]{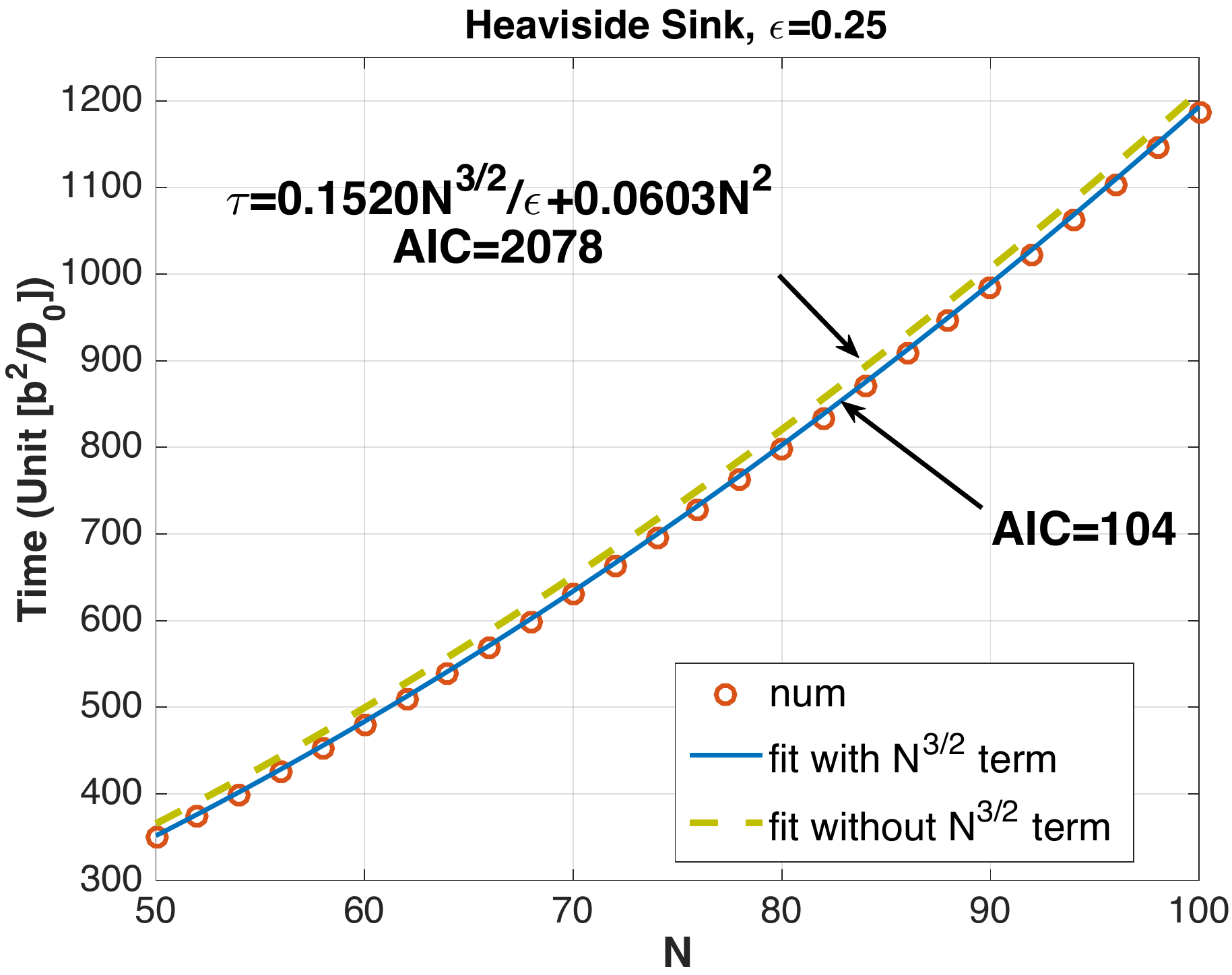}&{\bf (d)} \includegraphics[valign=t, scale=0.29]{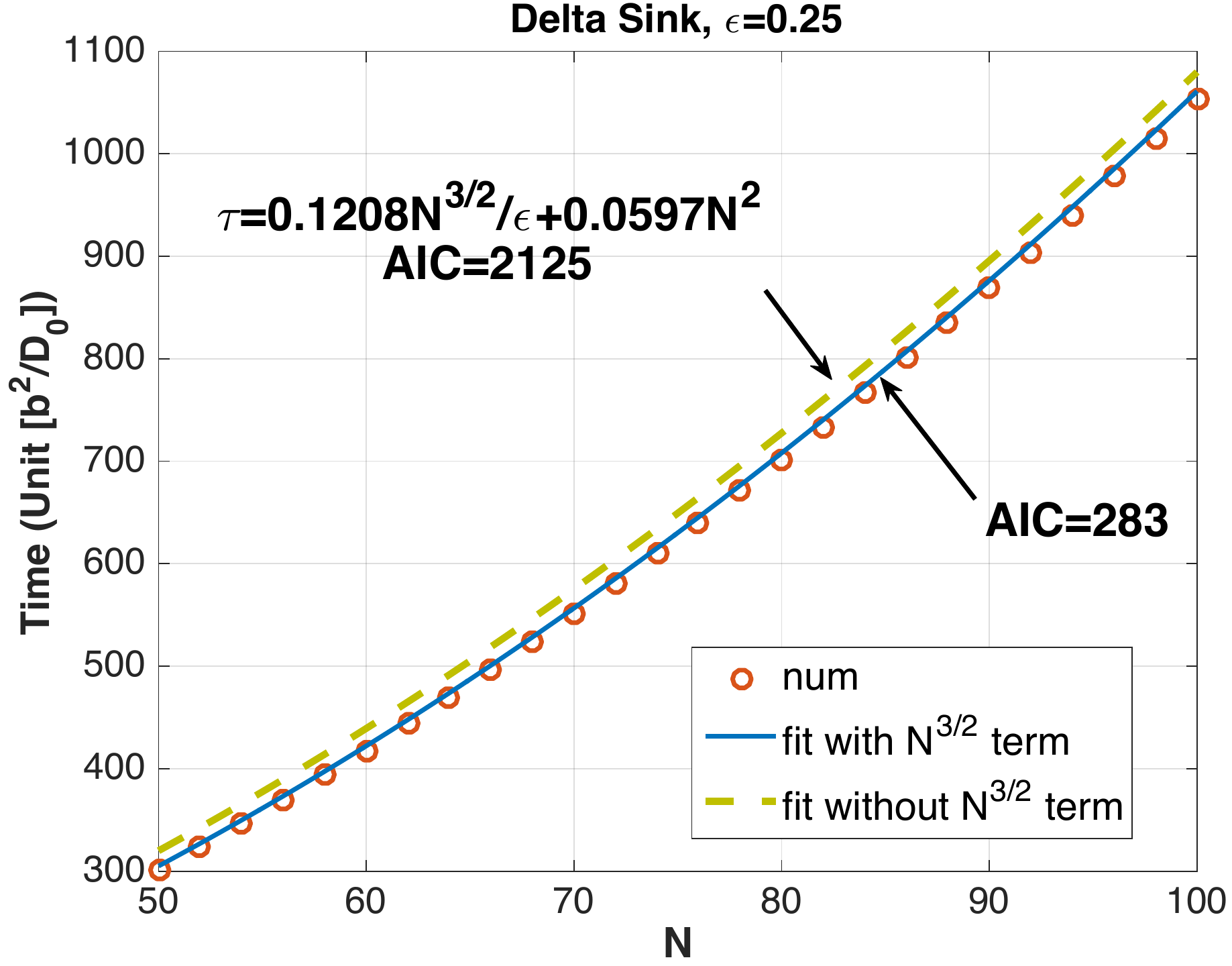}
\end{tabular}
\caption{(a,b) Comparison of results from regression fit and the WF theory. The fitted lines are plotted according to \cref{Heavi-fit} and \cref{DS-fit}. The WF results are obtained from numerical integration using \cref{Ct_heavi} and \cref{Ct_ds}. (c,d) Comparison of regression fit with and without $N\sqrt{N}$ term for $\epsilon=0.25$ and $50\le N\le 100$. }\label{comparison-2}
\end{figure}
With the scaling law at hand, we use the regression equation \cref{Heavi-fit} and \cref{DS-fit} to test other parameter ranges given by $\epsilon=0.25, 0.75$ and $50\le N\le 100, 150\le N \le200$ (see \Cref{comparison-2}). When $\epsilon=0.25$, which is relatively small, the fit agrees remarkably well with the WF approximation for both sinks and in both ranges of $N$. However, when $\epsilon=0.75$, the fit starts to deviate from the WF theory for both sinks and the difference grows with $N.$ Since we know that the WF theory can {\it overestimate} the looping time and the predictions from the regression fit are larger than those of the WF theory, we conclude that the predictions of the regression fit are not accurate for $\epsilon=0.75.$ For $\epsilon=0.25$, if one omits the new term $N\sqrt{N}$, \Cref{comparison-2} shows that for both sinks the regression lines deviate from the numerical integration points in this parameter range and also the values of the Akaike information criterion (AIC) are much higher. The AIC is a measure of the relative quality of a statistical model for a given set of data and and the lower AIC model is better \cite{Burnham1998}. This indicates that the $N\sqrt{N}$ term needs to be included in the model.
 
In the previous section, it was mentioned that WF approximates the denominator in the Laplace transform domain to get a semi-analytic form for the looping time and, as a result, the integrand is not a survival probability $\Sigma(t)$. However, the Volterra integral equation \cref{deconvolution} can be solved numerically by the trapezoidal method. Correspondingly, the survival probability is found numerically and each moment of passage time can be calculated. The most useful moment of passage time, of course, is the mean. 
 
 Rewrite \cref{deconvolution} to get
 \begin{equation}
 1=\int_0^t \Big( P_0q(t')\Big)\Big( \frac{C(t-t')}{C(\infty)}\Big) \rd t' 
 \end{equation}
 The kernel $C(t)/C(\infty)$ is the renormalized sink-sink correlation function and $P_0q(t)$ is the function to solve for. The survival probability is $\Sigma(t)=1-\int_0^t P_0q(t') \rd t'$. There is a difficulty because the kernel has a singular point at $t=0$ and the numerical integration is very stiff. With the perturbation result in \cref{perturbation}, we know the order of the singularity is 1/2, so careful handling of the kernel function at short time range $t$ is needed.

We plot in  \Cref{survival} the survival probability $\Sigma(t)$ predicted by the WF theory for two different values of $N$ and two different capture radius values $\epsilon$. As expected, the time integral of the survival probability recovers looping time from the WF theory in \cref{WF-appro}. In addition, we compare with the survival probability from the simulation data. This is computed by creating the histogram for the passage time and calculating the cumulative probability function for it. Then, the survival probability is one minus the cumulative probability function of the passage time. If the passage time follows a single exponential distribution, the survival probability will be an exponential function with the exponent given by the inverse of the looping time. 
For small capture radius $\epsilon=0.1,$ the survival probability function from WF agrees with the one from the simulation extremely well. It is interesting to note that the survival probability function obtained through simulation is an exponential function, even for large $N.$ This was also verified previously. The exponent of the exponential function is the inverse of the WF looping time.
However, for large capture radius $\epsilon=1,$ the survival probability function from the WF theory is clearly no longer an exponential function and it deviates from the simulation results which also do not show exponential behavior. 

\section{Discussion and future work}
We have reviewed the existing analytical theories for polymer looping in the Rouse model, including the SSS theory, the eigenvalue perturbation method, and the WF theory. 
Specifically, we reviewed the linkage between the J factor and the kinetics of looping by showing that J factor as the scaling of $N$ appears in the first order perturbation of the largest eigenvalue of the Fokker-Planck equation. 
We have rigorously investigated the conditions under which each asymptotic method holds and articulated the parameter range in which each of them is accurate. In particular, the main assumption for WF has been considerably relaxed, i.e, the solution has the space and time separation inside the sink
for the Heaviside sink and on the reactive surface for the delta sink.
 Our results show that the WF theory as a non-Markovian method can reproduce other existing analytical theories and it has the largest parameter range where it remains accurate. In fact, as far as the capture radius is small, i.e, $\epsilon$ is small,  the WF theory can predict the looping time extremely well. Moreover, under the same small $\epsilon$  condition, we are able to derive that the looping time follows the mixed scaling laws asymptotically and it is considered as the next order approximation of the Markovian approximation in powers of the capture radius $\epsilon$. Our findings have also been numerically verified.

We have to admit that further relaxing the time and space separation assumption systematically is not feasible analytically. In order to fully capture this non-Markovian effect, a considerably more involved method will be required. However, the WF theory as a general method of dimension reduction deserves more attention. It could be applied to
any reversible OU processes with complicated absorbing boundary conditions. The WF theory can also be used to extract semi-analytical or even analytical expressions for the mean first passage time and this may provide guidelines for numerical simulations and physical experiments. Natural extensions beyond the Rouse model include the hydrodynamical interaction or heterogeneous spring coefficients. Both models can be studied through independent but not identical normal modes with different drift and diffusion terms. Then the end-end correlation function $\phi(t)$ will be different, but the rest of the analysis carries through the same. 

\section*{Acknowledgments}
FY thanks particularly Attila Szabo (NIH) and 
Nathan Baker (PNNL) for helpful discussions and suggestions, and continuous encouragement in the initial phase of this project. The work of FY was partially supported by the U.S. Department of Energy Office of Science, Office of Advanced Scientific Computing Research, Applied Mathematics program, Collaboratory on Mathematics for Mesoscopic Modeling of Materials (CM4), under Award Number DE-SC0009280. The work of PS was partially supported by the Pacific Northwest National Laboratory Laboratory Directed Research and Development (LDRD) Project ``Multiscale modeling and uncertainty quantification for complex non-linear systems''.


\bibliographystyle{plain}
\bibliography{Master}

\begin{thebibliography}{10}

\bibitem{Amitai2012-Flexible}
A.~Amitai, I.~Kupka, and D.~Holcman.
\newblock Computation of the mean first-encounter time between the ends of a
  polymer chain.
\newblock {\em Phys. Rev. Lett.}, 109:108302, Sep 2012.

\bibitem{Benichou2005}
O.~B\`enichou, M.~Coppey, M.~Moreau, and G.~Oshanin.
\newblock Kinetics of diffusion-limited catalytically activated reactions: An
  extension of the {Wilemski-Fixman} approach.
\newblock {\em The Journal of Chemical Physics}, 123(19):194506, 2005.

\bibitem{Renewal2015}
O.~B\'enichou, T.~Gu\'erin, and R.~Voituriez.
\newblock Mean first-passage times in confined media: from markovian to
  non-markovian processes.
\newblock {\em Journal of Physics A: Mathematical and Theoretical},
  48(16):163001, 2015.

\bibitem{bressloff-rmp}
P.~C. Bressloff and J.~M. Newby.
\newblock Stochastic models of intracellular transport.
\newblock {\em Rev. Mod. Phys.}, 85:135--195, 2013.

\bibitem{Burnham1998}
K.P. Burnham and D.R. Anderson.
\newblock {\em Model Selection and Inference: A Practical Information-theoretic
  Approach}.
\newblock Intelligence, SS. of Lncs; 1501. Springer, 1998.

\bibitem{Chen2005}
Z.~Y. Chen, H.-K. Tsao, and Y.-J. Sheng.
\newblock Diffusion-controlled first contact of the ends of a polymer:
  Crossover between two scaling regimes.
\newblock {\em Phys. Rev. E}, 72:031804, Sep 2005.

\bibitem{ward2011}
A.~F. Cheviakov and M.~J. Ward.
\newblock Optimizing the principal eigenvalue of the laplacian in a sphere with
  interior traps.
\newblock {\em Mathematical and Computer Modeling}, 53(78):1394 -- 1409, 2011.
\newblock Mathematical Methods and Modeling of Biophysical Phenomena.

\bibitem{tomchou}
T.~Chou and M.~R. D'Orsogna.
\newblock First passage problems in biology.
\newblock In R.~Metzler, G.~Oshanin, and S.~Redner, editors, {\em First-Passage
  Phenomena and Their Applications}, pages 306--345. World Scientific
  Publisher, Singapore, 2014.

\bibitem{conover1999practical}
W.~J. Conover.
\newblock {\em Practical nonparametric statistics}.
\newblock Wiley series in probability and statistics: Applied probability and
  statistics. Wiley, 1999.

\bibitem{daoduc}
K.~Dao~Duc, Z.~Schuss, and D.~Holcman.
\newblock Oscillatory survival probability: Analytical, numerical study for
  oscillatory narrow escape and applications to neural network dynamics.
\newblock {\em SIAM Multiscale Modelling \& Simulation}, 2016.

\bibitem{Doi1975}
M.~Doi.
\newblock Diffusion-controlled reaction of polymers.
\newblock {\em Chemical Physics}, 9(3):455 -- 466, 1975.

\bibitem{DoiEdwards1988}
M.~Doi and S.F. Edwards.
\newblock {\em The Theory of Polymer Dynamics}.
\newblock International series of monographs on physics. Clarendon Press, 1988.

\bibitem{flory1989polymer}
P.J. Flory.
\newblock {\em Statistical Mechanics of Chain Molecules}.
\newblock Hanser Publishers, 1989.

\bibitem{Gardiner}
C.~Gardiner.
\newblock {\em Stochastic Methods: A Handbook for the Natural and Social
  Sciences}.
\newblock Springer Series in Synergetics. Springer-Verlag Berlin Heidelberg,
  2009.

\bibitem{Guerin2013}
T.~Gu\`erin, O.~B\`enichou, and R.~Voituriez.
\newblock Reactive conformations and non-markovian cyclization kinetics of a
  rouse polymer.
\newblock {\em The Journal of Chemical Physics}, 138(9):094908, 2013.

\bibitem{jacobson1950}
H.~Jacobson and W.~H. Stockmayer.
\newblock Intramolecular reaction in polycondensations. i. the theory of linear
  systems.
\newblock {\em The Journal of Chemical Physics}, 18(12):1600--1606, 1950.

\bibitem{VanKampen}
N.G.~Van Kampen.
\newblock {\em Stochastic Processes in Physics and Chemistry}.
\newblock North-Holland Personal Library. Elsevier, third edition, 2007.

\bibitem{kato-book}
T.~Kato.
\newblock {\em Perturbation Theory for Linear Operators}.
\newblock Classics in Mathematics. Springer-Verlag Berlin Heidelberg, 1995.

\bibitem{klapper1998}
I.~Klapper and H.~Qian.
\newblock Remarks on discrete and continuous large-scale models of {DNA}
  dynamics.
\newblock {\em Biophysical Journal}, 74(5):2504--2514, 1998.

\bibitem{levene2013}
S.~D. Levene, S.~M. Giovan, A.~Hanke, and M.~J. Shoura.
\newblock The thermodynamics of {DNA} loop formation, from j to z.
\newblock {\em Biochemical Society Transactions}, 41(2):513--518, 2013.

\bibitem{Likthman2006}
A.~E. Likthman and C.~M. Marques.
\newblock First-passage problem for the rouse polymer chain: an exact solution.
\newblock {\em Europhys. Lett.}, 75(6):971, 2006.

\bibitem{DNA-looping-review2}
K.S. Matthews.
\newblock {DNA} looping.
\newblock {\em Microbiological Reviews}, 56(1):123--136, 1992.

\bibitem{Szabo1996}
R.~W. Pastor, R.~Zwanzig, and A.~Szabo.
\newblock Diffusion limited first contact of the ends of a polymer: Comparison
  of theory with simulation.
\newblock {\em The Journal of Chemical Physics}, 105(9):3878--3882, 1996.

\bibitem{Portman2003}
J.~J. Portman.
\newblock Non-gaussian dynamics from a simulation of a short peptide: Loop
  closure rates and effective diffusion coefficients.
\newblock {\em The Journal of Chemical Physics}, 118(5):2381--2391, 2003.

\bibitem{nature1986}
M.~Ptashne.
\newblock Gene regulation by proteins acting nearby and at a distance.
\newblock {\em Nature}, 322(6081):697--701, 08 1986.

\bibitem{qian-jmb}
H.~Qian.
\newblock A mathematical analysis of the brownian dynamics of {DNA} tether.
\newblock {\em J. Math. Biol.}, 41:331--340, 2000.

\bibitem{Qian-OU}
H.~Qian.
\newblock Mathematical formalism for isothermal linear irreversibility.
\newblock {\em Proceedings of the Royal Society of London A: Mathematical,
  Physical and Engineering Sciences}, 457(2011):1645--1655, 2001.

\bibitem{rouse1953}
P.~E. Rouse.
\newblock A theory of the linear viscoelastic properties of dilute solutions of
  coiling polymers.
\newblock {\em The Journal of Chemical Physics}, 21:1272--1280, 1953.

\bibitem{DNA-looping-review1}
R.~Schleif.
\newblock {DNA} looping.
\newblock {\em Annual Review of Biochemistry}, 61(1):199--223, 1992.

\bibitem{shore-baldwin}
D.~Shore, J.~Langowski, and R.~L. Baldwin.
\newblock {DNA} flexibility studied by covalent closure of short fragments into
  circles.
\newblock {\em Proc. Natl. Acad. Sci. U.S.A.}, 78(8):4833--4837, 1981.

\bibitem{Sokolov2003}
I.~M. Sokolov.
\newblock Cyclization of a polymer: First-passage problem for a non-markovian
  process.
\newblock {\em Phys. Rev. Lett.}, 90:080601, Feb 2003.

\bibitem{Szabo1984}
A.~Szabo, G.~Lamm, and G.~H. Weiss.
\newblock Localized partial traps in diffusion processes and random walks.
\newblock {\em Journal of Statistical Physics}, 34(1):225--238, 1984.

\bibitem{Szabo1980}
A.~Szabo, K.~Schulten, and Z.~Schulten.
\newblock First passage time approach to diffusion controlled reactions.
\newblock {\em The Journal of Chemical Physics}, 72(8):4350--4357, 1980.

\bibitem{nature2016}
O.~B\'enichou T.~Gu\'erin, N.~Levernier and R.~Voituriez.
\newblock Mean first-passage times of non-markovian random walkers in
  confinement.
\newblock {\em Nature}, 534(7607):356, 2016.

\bibitem{peterthomas}
P.~J. Thomas.
\newblock A lower bound for the first passage time density of the
  suprathreshold ornstein-uhlenbeck process.
\newblock {\em J. Appl. Probability}, 48:420--434, 2011.

\bibitem{Toan2008}
N.~M. Toan, G.~Morrison, C.~Hyeon, and D.~Thirumalai.
\newblock Kinetics of loop formation in polymer chains.
\newblock {\em The Journal of Physical Chemistry B}, 112(19):6094--6106, 2008.

\bibitem{ward1993}
M.~J. Ward and J.~B. Keller.
\newblock Strong localized perturbations of eigenvalue problems.
\newblock {\em SIAM Journal on Applied Mathematics}, 53(3):770--798, 1993.

\bibitem{Weiss}
G.~H. Weiss.
\newblock A perturbation analysis of the wilemski�fixman approximation for
  diffusion controlled reactions.
\newblock {\em The Journal of Chemical Physics}, 80(6):2880--2887, 1984.

\bibitem{WilemskiFixman1}
G.~Wilemski and M.~Fixman.
\newblock Diffusion controlled intrachain reactions of polymers. i theory.
\newblock {\em The Journal of Chemical Physics}, 60(3):866--877, 1974.

\bibitem{WilemskiFixman2}
G.~Wilemski and M.~Fixman.
\newblock Diffusion controlled intrachain reactions of polymers. ii results for
  a pair of terminal reactive groups.
\newblock {\em The Journal of Chemical Physics}, 60(3):878--890, 1974.

\end{thebibliography}

\end{document}